\documentclass[trackchanges]{aastex631} % linenumbers
\usepackage{soul}

\submitjournal{ApJ}

\begin{document}

\title{Propagation of Interplanetary Shocks in the Inner Heliosphere\footnote{\today}}

%%%%%%%%%%%%%%%%%%%%%%%%%%%%%%%%%%%%%%%%%%%%%%%
%
% AUTHORS AND AFFILIATIONS
%
%%%%%%%%%%%%%%%%%%%%%%%%%%%%%%%%%%%%%%%%%%%%%%%

\author[0000-0003-3026-4456]{Munkhjargal~Lkhagvadorj}
\affiliation{Wigner Research Centre for Physics, Department of Space Physics and Space Technology \\
  Konkoly-Thege Mikl{\'o}s {\'u}t 29-33. \\
  H$-$1121~Budapest, Hungary}
\affiliation{Eötvös Loránd University, Faculty of Science \\
  P{\'a}zm{\'a}ny P{\'e}ter s{\'e}t{\'a}ny 1/A.\\
  H$-$1117~Budapest, Hungary}
%\affiliation{Institute of Physics and Technology, Department of Theoretical and High Energy Physics\\
%  Peace~Avenue~54b, Bayanzurkh~District\\
%ű  13330~Ulaanbaatar, Mongolia}

\author[0000-0001-9502-2816]{G{\'a}bor~Facsk{\'o}}
\affiliation{Wigner Research Centre for Physics, Department of Space Physics and Space Technology \\
  Konkoly-Thege Mikl{\'o}s {\'u}t 29-33. \\
 H$-$1121~Budapest, Hungary}
\affiliation{University of P{\'e}cs, Faculty of Sciences, Institute of Mathematics and Informatics\\
  Ifj{\'u}s{\'a}g {\'u}tja 6.\\
  H$-$7634 P{\'e}cs, Hungary}

\author[0000-0003-0845-0201]{Andrea~Opitz}
\affiliation{Wigner Research Centre for Physics, Department of Space Physics and Space Technology \\
  Konkoly-Thege Mikl{\'o}s {\'u}t 29-33. \\
  H$-$1121~Budapest, Hungary}

\author[0000-0001-6626-5253]{P{\'e}ter~Kov{\'a}cs}
\affiliation{Wigner Research Centre for Physics, Department of Space Physics and Space Technology \\
  Konkoly-Thege Mikl{\'o}s {\'u}t 29-33. \\
  H$-$1121~Budapest, Hungary}

\author[0000-0003-3240-7510]{David~G.~Sibeck}
\affiliation{NASA Goddard Space Flight Center \\
  8800 Greenbelt Rd, Greenbelt \\
  MD 20771, USA}

\correspondingauthor{G{\'a}bor Facsk{\'o}}
\email{facsko.gabor@wigner.hu, gabor.i.facsko@gmail.com}

\begin{abstract}
  Interplanetary shocks are one of the crucial dynamic phenomena in the Heliosphere. They can accelerate particles to high energies, generate plasma waves, and can trigger geomagnetic storms in Earth's magnetosphere, significantly impacting technological infrastructure. In this study, two IP shock events are selected to study the temporal variations of the shock parameters using magnetometer and ion plasma measurements of the STEREO$-$A and B, the Wind, Cluster fleet, and the ACE spacecraft. The shock normal vectors are determined using the minimum variance analysis (MVA) and the magnetic coplanarity methods.  During the May 7, 2007, event, the shock parameters and the shock normal direction remain consistent across each spacecraft crossing of the shock. The shock surface appears to be tilted almost the same degree as the Parker spiral, and the driver could be a Co--Rotating Interaction Region (CIR). During the April 23, 2007 event, the shock parameters do not change significantly except for the shock angle $\theta_{Bn}$, however, the shape of the IP shock appears to be twisted along the perpendicular direction to the Sun-Earth line as well. The driver of this rippled shock is Stream Interaction Region (SIR)/CIR as well.
\end{abstract}


% Main body

\section{Introduction} \label{sec:intro}
The solar corona is hotter than the photosphere, the chromosphere, and the transient layers beneath it. As a result, the high temperatures ionize atoms, creating a plasma of free-moving electrons and ions, the so-called solar wind. Historically, \citet{parker58:_dynam_inter_gas_magnet_field} predicted the existence of the solar wind and coined terms describing it. He deduced it based on German astronomer Ludwig Bierman's observation of how the comet tail always points away from the Sun \citep{biermann57:_solar}. The existence of the solar wind was confirmed by the Mariner 2 spacecraft \citep{snyder65:_inter_solar_wind_measur_marin_ii}. The solar wind is a collisionless plasma, whose speed is both supersonic and super-Alfvenic. The latter means that its speed exceeds the Alfv{\'e}n speed, which is a representative speed of magnetohydrodynamic waves in a plasma. Due to the abrupt change of the solar wind speed from supersonic to subsonic (e.g. by encountering an obstacle), a shock wave emerges. Interplanetary (IP) shock waves are common in the heliosphere, which is a bubble-like region of space surrounding the Sun and extending far beyond the orbits of the planets and is filled with the solar wind. There are many varieties of shocks such as planetary bow shocks, shocks that rise due to the stream interaction regions \citep[SIR; see for example in][]{richardson18:_solar,koban23:_orien_cirs,silwal24:_multis_energ_partic_accel_assoc}, and coronal mass ejection (CME) driven shocks. After completing a solar rotation, SIRs are referred to as Corotating Interaction Regions (CIRs). IP shocks are one of the main and efficient accelerators of energetic particles in the heliosphere \citep{tsurutani85:_accel_au, keith21:_histor}. These accelerated particles can enter the geomagnetic field and are hazardous to astronauts and satellites. IP shocks driven by CMEs can trigger large geomagnetic storms that can damage oil and gas pipelines and interfere with electrical power infrastructures \citep{gonzalez94:_what}. GPS navigation and high-frequency radio communications are also affected by ionosphere irregularities brought on by geomagnetic storms \citep{cid14} and can cause internet disruptions around the world \citep{jyothi21:_solar_super,facsko24:_space_weath_effec_critic_infras}. Therefore, IP shocks are important in determining and understanding space weather.

Some dayside magnetospheric transient events \citep[see references therein]{zhang22:_daysid_trans_phenom_their_impac_magnet_ionos} are generated by the interaction of a discontinuity frozen in the solar wind and the terrestrial bow shock. This fact raises the questions of how stable the discontinuities are, how long these phenomena travel with the solar wind, and whether they evolve during their travel, or show some wave dispersion \citep[Istv{\'a}n Ballai, personal communication]{facsko05:_ident_statis_analy_hot_flow}. Dayside magnetospheric transient events are generated by tangential (TD) and rotational discontinuities (RD). However, there are catalogues for neither TDs nor RDs. Therefore, it is easier to study IP shocks because many were observed and listed previously (see Section~\ref{sec:ipshock}).

\citet{reshetnyk10:_geomet_proper_solar_wind_discon_earth_orbit} studied the 2D curvature of two IP shock events. \citet{reshetnyk10:_geomet_proper_solar_wind_discon_earth_orbit}'s paper showed that we do not know so much about the shape of discontinuities, especially about those of IP shocks. \citet{facsko10:_study_stereo_ulyss} extended the previous study to 3D. More events and more discontinuities (IP shocks) were needed. A special automated shock identification method was developed \citep{facsko24:_bow}, however, finally, the appropriate events to study the temporal development and 3D shape of IP shocks were found in an open database developed by \citet{nikbakhsh14:_inter,kilpua15:_proper_earth}. Finally, the origin of the IP shocks is also very interesting. Based on our current knowledge the shock could be CME driven, the result of SIR/CIR development, or caused by a flare. The determination of the 3D shape of IP shocks with a backward projection might explain the origin of these very important events.

The shape of an IP shock front depends on the inhomogeneities of the solar wind through which the shock is propagating \citep{heinemann74:_shapes,odstrcil99:_three,dryer04:_real_hallow}. \citet{russell83:_multipa,russell83:_multipb} studied IP shock using two- and four-spacecraft measurements assuming that the IP shocks were planar locally. \citet{terasawa05:_deter_octob} observed $\mathrm{15-20}^{\circ}$ shock surface ripple using Geotail \citep{nishida94:_geotail} and Advanced Composition Explorer \citep[ACE;][]{stone98:_advan_compos_explor} data. \citet{szabo05:_multi_spacec_obser_inter_shock} reported discrepancies of IP shock normal directions observing the same IP shock using ACE and Wind  \citep{lin95:_three_dimen_plasm_energ_partic,wilson21:_quart_centur_wind_spacec_discov}, or Wind and IMP--8  \citep{team74:_explor} spacecraft. \citet{lepping03:_two_mhd} determined IP shock normals using Wind and IMP--8 observations at 1\,AU (Astronomical~Unit~$\approx 1.5\times 10^{8}\,\mathrm{km}$). The average radius of the IP shock curvature was a few million\,km. \citet{neugebauer05:_multis} collected and analyzed 26 IP shocks between the Earth's bow shock and $L_{1}$ Lagrangian point. The average radius of the IP shock curvatures was a bit larger than \citet{lepping03:_two_mhd}'s result, however, it also was a few million\,km. This result agreed with previous simulation results \citep{neugebauer05:_multis}.

\citet{lario08:_influen} observed the same IP shock at 1\,AU using ACE measurements and at 5.4\,AU using Ulyesses observations. However, \citet{lario08:_influen} studied the solar energetic particle (SEP) events. The interplanetary solar wind properties changed the properties of the SEP events in this study.

The Helios probe \citep{howard75:_projec_helios} observed IP shocks to evolve and change their behaviour at various heliospheric distances in the inner heliosphere \citep{lai12:_radial_variat_inter_shock_inner_helios} and between 1 and 5\,AU by the Ulysses mission \citep{wenzel92:_ulyss_mission,burton92:_ulyss,zhao18:_unusual_energ_partic_flux_enhan}. \citet{trotta23:_multi} studied shocklets using ACE, Wind, DSCOVR (Deep Space Climate Observatory), Time History of Events and Macroscale Interactions during Substorms \citep[THEMIS,][]{angelopoulos08:_themis_mission} B and C measurements. \citet{trotta24:_proper_inter_shock_obser} observed a CME driven shock at 0.07\,AU and 0.7\,AU using the Parker Solar Probe \citep[PSP,][]{fox16:_solar_probe_plus_mission}, and the Solar Orbiter \citep[SolO,][]{marsch05:_solar_orbit} observations. The PSP crossed a thick magnetosheath at the flank according to the early evolution phase of the CME. The SolO observed high energy (2\,MeV) shock accelerated protons and a much more disturbed shock environment than the one observed by PSP. Therefore, the local small-scale environment of the shock was very different at different locations of the heliosphere.

In this study, we focus on the macroscopic structure and parameters of the observed IP shock events. The structure of this paper is as follows: we first list and describe briefly the missions, instruments, and databases that we used in this study in Section~\ref{sec:mis}. We present the analyzed IP shock events in Section~\ref{sec:ipshock}, then we describe the applied analysis methods in Section~\ref{sec:methods}. Then we introduce the datasets that we analysed in Section~\ref{sec:obs}.  We discuss and analyse the observations in Section~\ref{sec:disc}. Finally, we summarize the results of our study in Section~\ref{sec:sum}.

\section{Missions and instruments} \label{sec:mis}

In this paper, Solar Terrestrial Relations Observatory \citep[STEREO;][]{kaiser08:_stereo_mission} A and B, Wind \citep{lin95:_three_dimen_plasm_energ_partic,wilson21:_quart_centur_wind_spacec_discov}, ACE, and the Cluster \citep{credland97:_clust_mission,escoubet01:_introd} spacecraft magnetic field, ion plasma, and spacecraft potential data are used.

\subsection{The STEREO mission} \label{sec:stereo}

NASA's twin STEREO A and B spacecraft were launched on October 26, 2006, from Kennedy Space Center. In heliospheric orbit at 1\,AU, STEREO$-$A (Ahead) leads while STEREO$-$B (Behind) trails Earth. The two spacecraft separate at $44^{\circ}$ in heliographic longitude from each other annually. Both spacecraft were equipped with two instruments and two instrument suites, with a total of 13 instruments on each spacecraft.

The PLAsma and SupraThermal Ion Composition (PLASTIC) instrument measures protons and alpha particle parameters, as well as the composition of heavy ions in the solar wind plasma \citep{galvin08:_plasm_suprat_ion_compos_plast}. The In-situ Measurements of Particles and CME Transients magnetic field experiment (IMPACT) suite of instruments consists of seven instruments among which three are located on a 6-meter long boom. The other four are installed in the main part of the spacecraft \citep{acuna08:_stereo_impac_magnet_field_exper}. The IMPACT measures the parameters of protons, heavy ions, and electrons, and the MAG magnetometer sensor in it measures the in situ magnetic fields in a range of $\pm 512\,nT$ with 0.1\,nT resolution \citep{kaiser07:_stereo}.

We downloaded all data from NASA's Coordinated Data Analysis Web (CDAWeb, \url{https://cdaweb.gsfc.nasa.gov/}). We obtained the STEREO$-$A and B magnetic field observation and the ion plasma data from the STEREO IMPACT with time resolution of 100\,ms and the STEREO PLASTIC instruments with time resolution of 1\,minute, respectively. The magnetic field and the plasma data of the STEREO$-$A and B are in the RTN spacecraft coordinate system, where R is radially outward from the Sun, T is along the planetary orbital, and N is northward direction.

\subsection{The Wind mission} \label{sec:wind}

NASA's Wind spacecraft was launched on November 1, 1994, \citep{lin95:_three_dimen_plasm_energ_partic,wilson21:_quart_centur_wind_spacec_discov}. The Wind was initially sent to $L_1$ Lagrange point, however, the reaching of $L_{1}$ was delayed to be able to study in advance the terrestrial magnetosphere and lunar environment, as well. Following a sequence of orbital adjustments, the Wind spacecraft was positioned in a Lissajous orbit close to the $L_1$ Lagrange point in early 2004 for studying the incoming solar wind \citep{team20:_wind_spacec}.

The spacecraft is equipped with eight instruments, however, we used only the Magnetic Field Investigation \citep[MFI;][]{lepping95:_wind_magnet_field_inves} and the Three-Dimensional Plasma and Energetic Particle Investigation \citep[3DP][]{lin95:_three_dimen_plasm_energ_partic} measurements in this study. The MFI consists of two magnetometers at the 12-meter boom, they measure magnetic field from 4$\,$nT{,} to 65536$\,$nT with a time resolution of 22 or 11 vectors per second in the calibrated high-resolution mode, while in the primary science mode the resolution can be three seconds, one minute and one hour \citep{team22:_wind_data_sourc}. The 3DP measures the solar wind key parameters such as velocity, density, and temperature.

From CDAWeb, we got Wind magnetic and plasma data from the MFI and 3DP instruments with a time resolution of 3\,s, for both. The magnetic field and the plasma data of the Wind are in the Geocentric Solar Ecliptic System (GSE) coordinate system, where the X-axis is pointing to the Sun from the Earth, Y-axis is in the ecliptic plane showing opposite to the planetary motion, and Z-axis is the northward direction.

\subsection{The ACE mission} \label{sec:ace}

NASA's ACE spacecraft was launched on August 25, 1997, \citep{stone98:_advan_compos_explor}. The spacecraft is located at $L_{1}$ Lagrangian point similarly to the Wind spacecraft. The spacecraft is equipped with nine primary scientific instruments and one engineering instrument. We used in the study the measurements of the Magnetometer \citep[MAG;][]{smith98:_ace_magnet_field_exper}, and the Solar Wind Electron, Proton and Alpha Monitor \citep[SWEPAM;][]{mccomas98:_solar_wind_elect_proton_alpha}. The MAG consists of twin triaxial flux-gate magnetometers such that magnetometer sensors have between 3 and 6 vectors $s^{-1}$ resolutions for continuous observation of the interplanetary magnetic field \citep{smith98:_ace_magnet_field_exper}.

From CDAWeb, we downloaded the ACE magnetic and plasma data from the ACE MAG with a time resolution of 1\,s and from SWEPAM with a time resolution of 64\,s, respectively. The magnetic field and the plasma data are both in the RTN and GSE coordinate systems.

\subsection{The Cluster fleet} \label{sec:cluster}

ESA's Cluster constellations consist of four satellites, which were launched on 16 July and 9 August 2000 \citep{credland97:_clust_mission,escoubet01:_introd}. The Cluster satellites orbit in a tetrahedral formation around Earth. The perigee and apogee of the orbit are approximately four and 19.6\,$R_{E}$, respectively \citep{zhang10:_emic_he}. Each of the four satellites is equipped with 11 identical instruments. We used the data of the Cluster Ion Composition \citep[CIS;][]{reme97:_clust_ion_spect_exper,reme01:_first_earth_clust_cis}, the Fluxgate Magnetometer \citep[FGM;][]{balogh97:_clust_magnet_field_inves,balogh01:_clust_magnet_field_inves}, and the Electric field and waves \citep[EFW;][]{gustafsson97:_elect_field_wave_exper_clust_mission,gustafsson01:_first_clust_efw} instruments in this study .

The FGM is composed of two tri-axial fluxgate magnetometers, which are installed on one of the two 5-meter radial booms. It measures in the dynamic range of $\pm 65,536$\,nT. At the highest dynamic level, the resolution is $\pm8$\,nT, and the time resolution is 100 vectors per second \citep{balogh97:_clust_magnet_field_inves,balogh01:_clust_magnet_field_inves,organization21:_fluxg_magnet_fgm}. The CIS instrument measures three-dimensional ion distribution, and it is composed of two distinct sensors: the Composition Distribution Function (CODIF) sensor and the Hot Ion Analyzer (HIA) sensor \citep{reme97:_clust_ion_spect_exper,reme01:_first_earth_clust_cis}. The CIS experiment is not operational for Cluster-2 and the HIA sensor is switched off for Cluster-4 due to a problem with the high voltage of the electrostatic analyzer \citep{team21:_clust_activ_archiv}. Hence, for Cluster-2 and Cluster-4, the EFW instrument measurement became more indispensable for the analyses because we could obtain electron density in another way, independently from ion and electron plasma measurements \citep{gustafsson97:_elect_field_wave_exper_clust_mission,gustafsson01:_first_clust_efw}.

As the spacecraft travels through the plasma environment, it acquires an electric charge as a result of contact with charged particles. This charging process results in an electrical potential discrepancy between the spacecraft and the plasma around it, a phenomenon referred to as spacecraft potential. The EFW instrument measures the spacecraft potential and the electron density of the plasma could be calculated using an empirical formula.
\begin{equation}\label{eq:empdens}
N_{e}=200(V_{sc})^{-1.85},
\end{equation}
where $N_{e}$ is the calculated electron density and $V_{sc}$ is the Cluster EFW spacecraft potential \citep{sandhu16:_clust,trotignon10:_whisp_relax_sound_clust_activ_archiv,trotignon11:_calib_repor_whisp_measur_clust}.

From CDAWeb, we acquired the magnetic data of Cluster satellites from the Cluster FGM with a time resolution of 4\,s for all the four Cluster satellites, ion data from the Cluster CIS with a time resolution of 4\,s for the Cluster~SC1 and SC3 satellites, and the spacecraft potential data from EFW with a time resolution of 4\,s for the Cluster~SC2 and SC4 satellites where CIS measurements were not available. All the Cluster data were in the GSE coordinate system.

\section{IP shock events and transformations} \label{sec:ipshock}

IP shock candidates are chosen from the shock lists in the Database of Heliospheric Shock Waves maintained at the University of Helsinki \url{http://www.ipshocks.fi/}. For the event selection, we chose the year 2007 because STEREO$-$A and STEREO$-$B were close to each other near the Sun-Earth line. In the first event, May 07 2007, the selected spacecraft are STEREO$-$A, STEREO$-$B, Wind, and the four Cluster satellites. For the second event, April 23, 2007, the spacecraft are STEREO$-$A and B, ACE, and Wind.

After choosing the shock candidates and obtaining the data, we transformed data coordinates in such a way that all the coordinate systems are changed to the Heliocentric Earth Ecliptic (HEE) coordinate system, where the X-axis is toward the Earth from the Sun, Z-axis is the ecliptic northward direction, while Y axis completes the right-handed Cartesian system. To transform from the RTN and the GSE coordinate systems to the HEE coordinate system, we used Transformation de REp{\`e}res en Physique Spatiale (TREPS; \url{http://treps.irap.omp.eu/}) online tool. The TREPS tool, which is developed by the French Plasma Physics Data Centre (CDPP), the national data centre in France for the solar system plasmas, is based on SPICE (Spacecraft, Planet, Instrument, C-matrix, and Events) information system kernels created by National Aeronautics and Space Administration (NASA)/Navigation and Ancillary Information Facility (NAIF) tool \citep{genot18:_treps}.

\section{Analysis methods} \label{sec:methods}

Usually coplanarity method is applied to find the shock normal. The coplanarity method is based on the magnetic coplanarity theorem, which states that the magnetic field vectors on both sides of the shock and the shock normal lie in the same plane. Similarly, the velocity on both sides or, in other words, the velocity jump through the shock also lie in the same plane. The method of magnetic coplanarity is straightforward to implement, it requires only magnetic field data for applying the method \citep{paschmann98:_analy_method_multi_spacec_data,schwartz98:_shock_discon_normal_mach_number_relat_param}.

We select two intervals upstream and downstream of the shock, calculate average upstream and downstream magnetic field vectors using the values of the interval, and determine the shock normal vector using the coplanarity method. However, in this case, the magnetic field varies strongly, therefore the calculated shock normal vector looks very sensitive for the selected intervals if one uses only the coplanarity method. Hence, the calculated shock normals for each spacecraft pointed in different directions. The result cannot be explained physically, therefore applying the coplanarity method only is pointless in this case. Hence, we need the support of another method in this case.

Therefore, we use Minimum Variance Analysis (MVA) \citep{sonnerup98:_minim_maxim_varian_analy, schwartz98:_shock_discon_normal_mach_number_relat_param} and magnetic co-planarity (CP) methods to determine the IP shock normals. The MVA technique is based on the assumption that variations in the magnetic field would be observed when a single spacecraft passes through a 1D (in reality, it is a 2D or 3D transition layer) current layer or wavefront. The divergence of the magnetic field is constrained $\left(\nabla \cdot \mathbf{B} = 0\right)$, therefore the normal component of the magnetic field must remain constant. If such a normal direction can be found, then the variations in the magnetic field are zero or at least have a minimum variance \citep{sonnerup67:_magnet_struc_attit_explor_obser,sonnerup98:_minim_maxim_varian_analy,wang24:_solar_wind_curren_sheet}. MVA can only work if there are less fluctuations in the direction normal to the shock than in the direction perpendicular to magnetic fields upstream and downstream from the shock.

By comparing the two methods, the upstream and downstream time intervals of the magnetic field measurements are set. We set up the following requirements to accept the results:
\begin{enumerate}
\item The angle between the shock normal vectors obtained by the minimum variance analysis (MVA) and the magnetic coplanarity method must be less than $15^{\circ}$.
\item The ratio between the intermediate eigenvalue and the smallest eigenvalue should be greater than 2 in the MVA method.
\end{enumerate}
The above methods have been used in previous studies for shocks \citep{bebesi10:_slow_jovian,bebesi19:_obser_kronian,borodkova19:_fine_struc_inter_shock_front,nikbakhsh14:_inter,shan13:_compar_mva_venus,terasawa05:_deter_octob}. Similar conditions have been used in other studies for tangential discontinuities \citep{facsko08:_clust,facsko09:_clust,facsko10:_study_clust}. % sonnerup67:_magnet_struc_attit_explor_obser

We apply both methods for every single spacecraft. The shock normal vectors calculated independently point to a similar direction. This result is acceptable and could be explained physically. We suppose that the same shock passes all spacecraft at different times and locations. The shock is supposed to be a bit curved and perhaps rippled. Therefore, applying both methods leads to a valuable and acceptable result.

\section{Observations} \label{sec:obs}

With the visual inspection of the magnetic field and plasma data, we identified jumps in magnetic field ($\mathbf{B}$), speed ($\mathbf{V}$), density (N), and temperature (T). Using these quantities, we determined the times when shock events occurred for each spacecraft data set. The common origin of a shock event detected by different spacecraft can be concluded from the similar time profiles of the magnetic field and plasma data recorded in each spacecraft position near the shock. From the positions of the spacecraft and the times of shock arrivals to the different positions, the shock geometry and the propagation parameters (velocity, direction) can be deduced. The following case studies of two events are discussed in reversed time order. First, we analyzed the shock event on May 7, 2007, and then the event on April 23, 2007.

\subsection{Event May 7, 2007} \label{sec:event20070507}

\begin{figure*}
 \centering
 \includegraphics[angle=0,width=0.99\textwidth]{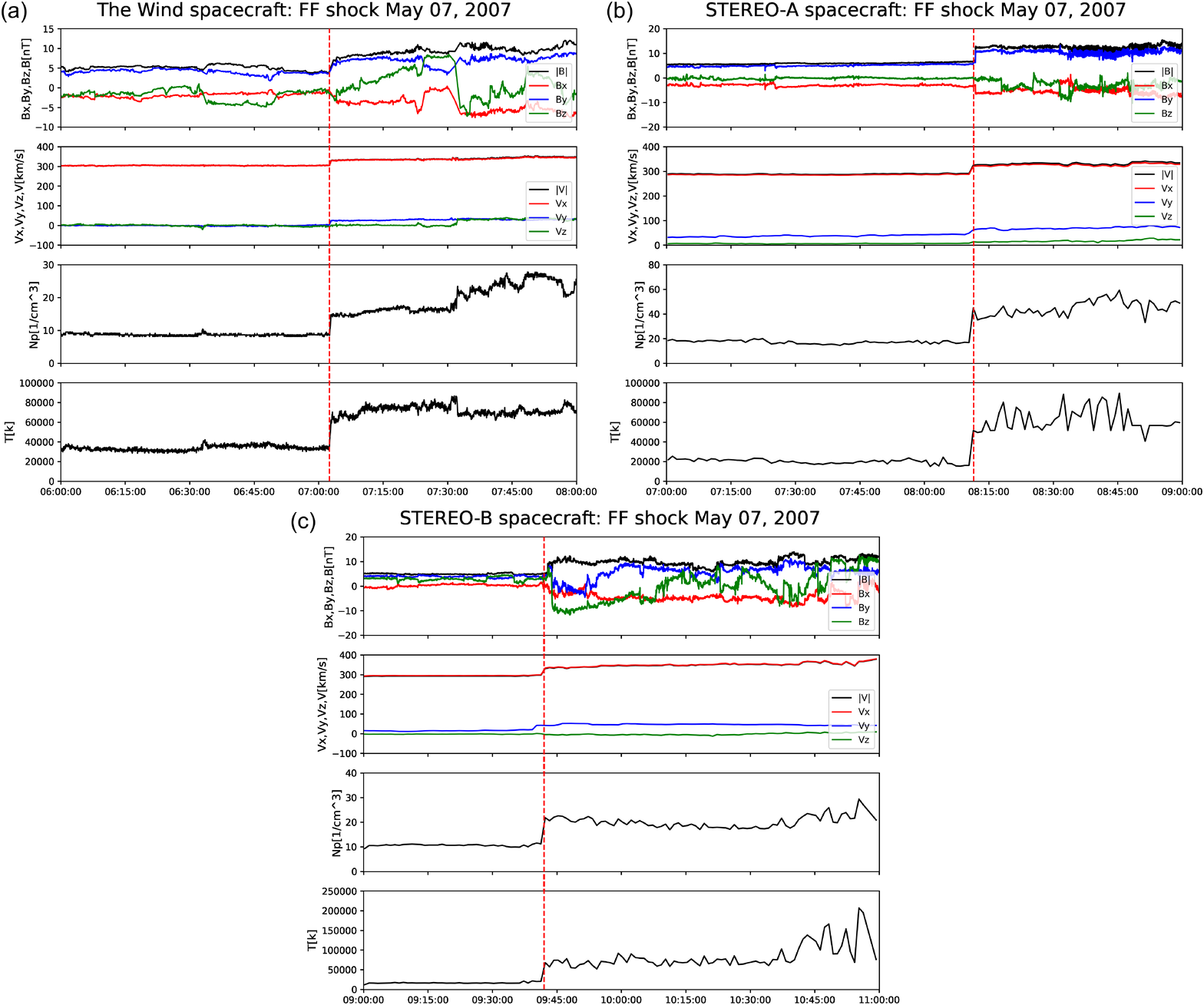}
\caption{The panels show from top to bottom, the magnetic field magnitude as well as its components, the total velocity, and its components, density, and temperature. The dashed red line represents the exact shock time. The duration of the plots is two hours. (a) The plot of the shock detected by the Wind spacecraft on May 7, 2007, at 07:02:30 (UTC). FF stands for the fast forward shock, which means the shock is travelling away from its driver. (b) The plot of the shock detected by the STEREO$-$A spacecraft on May 7, 2007, at 08:11:30 (UTC). (c) The plot of the shock detected by the STEREO$-$B spacecraft on May 7, 2007, at 09:42:00 (UTC). \label{fig:ip0507}}
\end{figure*}

\begin{figure*}
\centering
\includegraphics[width=0.99\textwidth]{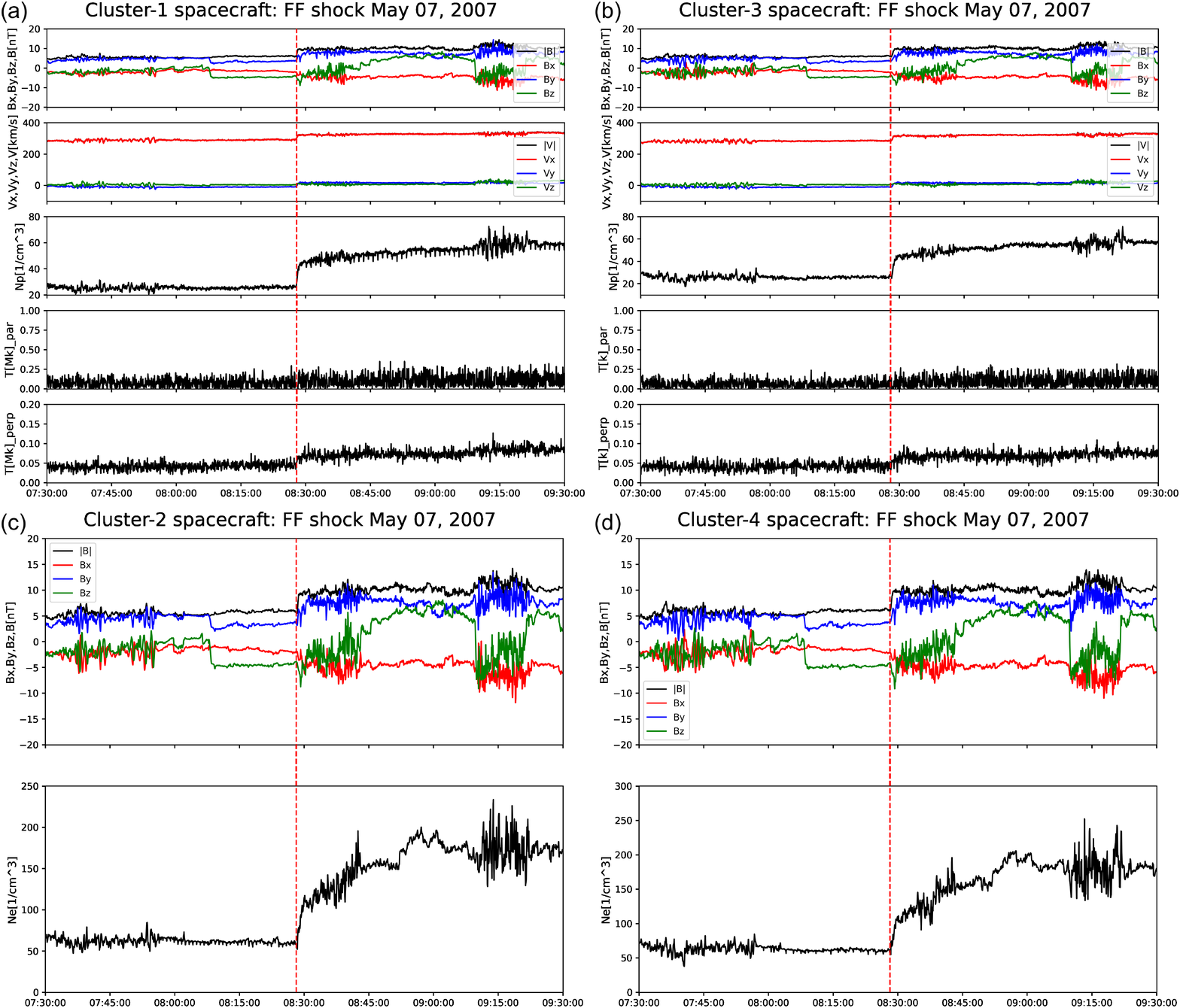}
\caption{The plot of the shock detected by the Cluster SC1 (a), SC2 (b), SC3 (c), and SC4 (d) on May 7, 2007, at 08:27:55 (UTC). The panels show from top to bottom, the magnetic field magnitude as well as its components, the total velocity, and its components, density, and parallel and perpendicular temperatures. The dashed red line represents the exact shock time. The duration of the plot is two hours. Only the magnetic field magnitude and density are shown for Cluster SC2 and SC4. \label{fig:cl0507}}
\end{figure*}

On May 7, 2007, the Wind spacecraft detected the shock first at 07:02:30 (UTC). After that, the STEREO$-$A spacecraft detected the shock at 08:11:30 (UTC), and then the STEREO$-$B spacecraft detected the shock at 09:42:00 (UTC). The four Cluster satellites detected the shock as well. Cluster SC1 detected the shock at 08:27:55 (UTC), and Cluster SC3 detected the shock at 08:28:00 (UTC). The magnetic field and the plasma parameters are shown in Figure~\ref{fig:ip0507} for the Wind (a), for the STEREO$-$A(b), for the STEREO$-$B (c), in Figure~\ref{fig:cl0507} for the Cluster SC1 (a), and SC3 (b). There is no indication in the shock database that Cluster SC2 and SC4 detected a shock in the analysed period. However, since the four Cluster spacecraft are located close to each other, they all had to encounter the given shock. Without plasma data on Cluster SC2 and SC4 satellites, a shock cannot be confirmed based solely on the magnetic field data. However, using the empirical formula described in Section~\ref{sec:cluster} the electron density could be estimated using EFW spacecraft potential data. Therefore, comparing the magnetic field and the electron plasma density profile, the shock is determined. The Cluster SC2 detected the shock at 08:28:10 (UTC), according to the magnetic field and density plots (Figure~\ref{fig:cl0507}c). The Cluster SC4 detected the shock at the same time as SC2 (Figure~\ref{fig:cl0507}d). Similarly, The electron density parameter is obtained by the empirical formula (Eq.~\ref{eq:empdens}).

\begin{figure*}
\centering
\includegraphics[width=0.99\textwidth]{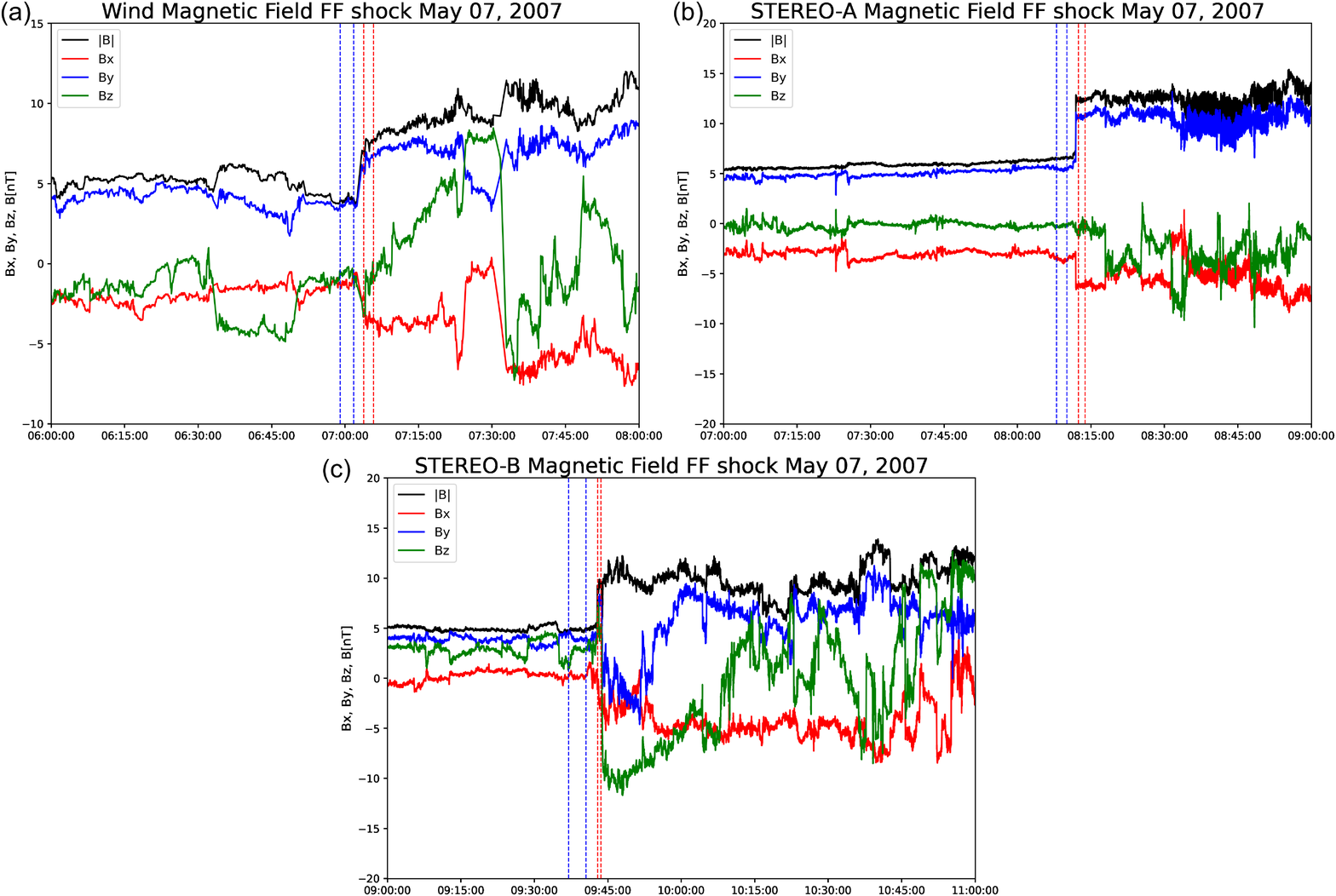}
\caption{(a) The Wind magnetic field measurements on May 7, 2007. The upstream $\Delta t_{up}$ is between (06:59:00 - 07:01:47), and the downstream $\Delta t_{down}$ is between (07:03:50 - 07:05:50). (b) The STEREO$-$A magnetic field measurements on May 7, 2007. The upstream $\Delta t_{up}$ is between (08:08:00 - 08:10:05), and the downstream $\Delta t_{down}$ is between (08:12:30 - 08:13:50). (c) The STEREO$-$B magnetic field measurements on May 7, 2007. The upstream $\Delta t_{up}$ is between (09:37:00 - 09:40:35), and the downstream $\Delta t_{down}$ is between (09:42:57 - 09:43:38). FF stands for the fast forward shock, which means the shock is travelling away from its driver. The chosen intervals of the upstream and downstream magnetic fields were used for minimum variance analysis and co-planarity methods. The blue and red dashed lines each represent the upstream starting time and ending time and the downstream starting time and ending time, respectively. \label{fig:b20070507}}
\end{figure*}

\begin{figure*}
\centering
\includegraphics[width=0.99\textwidth]{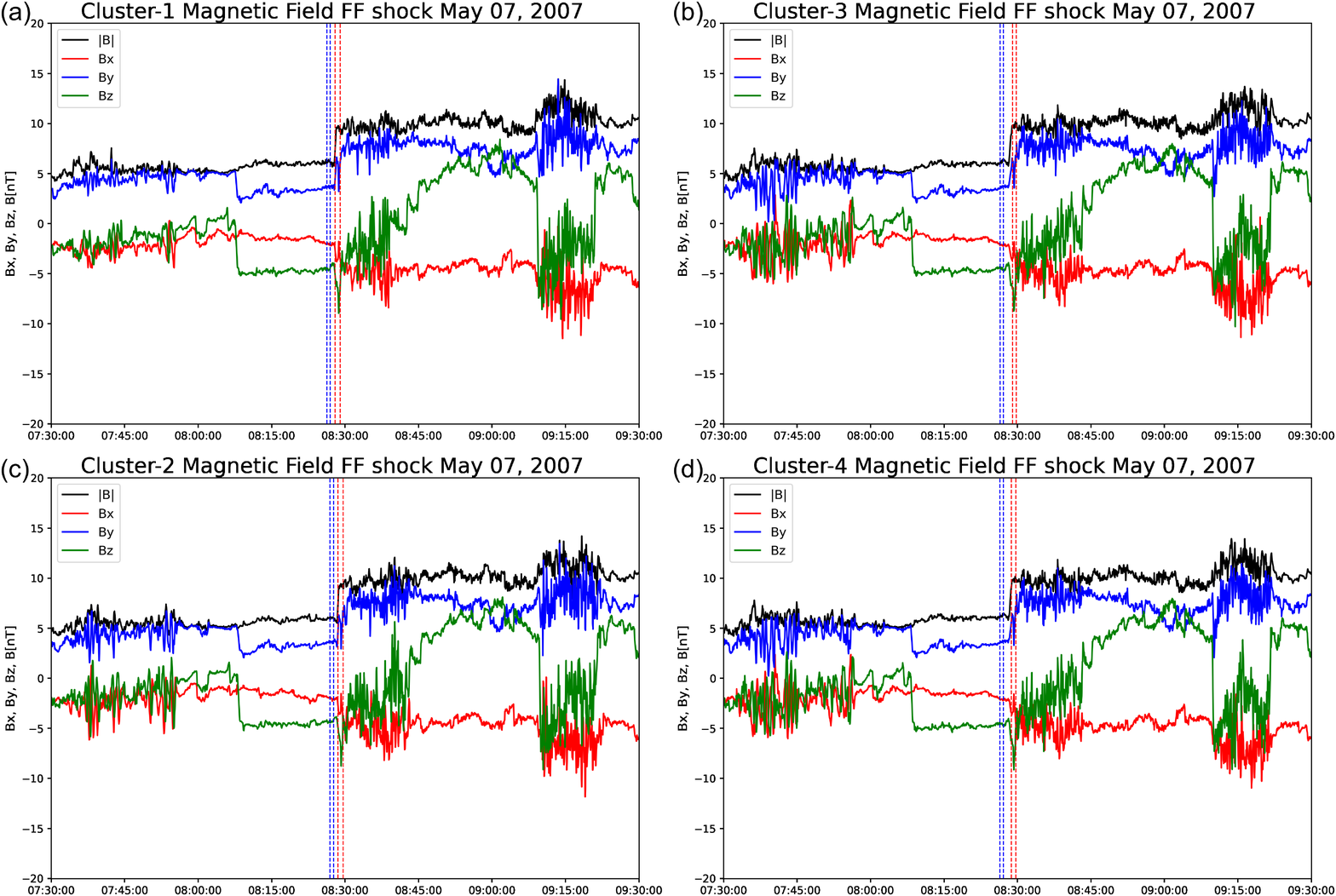}
\caption{The Cluster SC1 (a), SC3 (b), SC2 (c), and SC4 (d), respectively. B magnetic field measurements on May 7, 2007. The upstream $\Delta t_{up}$ is between (08:26:10 - 08:27:00), (08:26:30 - 08:27:10), (08:26:56 - 08:27:40), (08:26:25 - 08:27:10) and the downstream $\Delta t_{down}$ is between (08:28:00 - 08:29:00), (08:29:00 - 08:29:47), (08:28:32 - 08:29:37), (08:28:32 - 08:29:37), (08:28:45 - 08:29:40) for Cluster SC1, SC2, SC3, SC4, respectively. The symbols and details of the figures are the same as \ref{fig:b20070507}. The chosen intervals of the upstream and downstream magnetic fields were used for minimum variance analysis and co-planarity methods. \label{fig:clb20070507}}
\end{figure*}

Here we list all magnetic field observations of the spacecraft for the event and apply the MVA and the CP methods to them. The upstream and downstream time intervals that most agree between the MVA and the CP for all the spacecraft are shown in Table~\ref{tab:timeintervalsmay}, and their corresponding magnetic field plots are shown in Figure~\ref{fig:b20070507} for Wind ({a), STEREO$-$A (b), STEREO$-$B (c), while in Figure}~\ref{fig:clb20070507} for Cluster spacecraft. Table~\ref{tab:timeintervalsmay} also shows the ratio between the smallest eigenvalue $\lambda_{3}$ and the intermediate eigenvalue $\lambda_{2}$ as well as the angle between the MVA normal and CP normals in the determined upstream and downstream time intervals for each spacecraft.

The accepted upstream and downstream time intervals were properly selected considering the angle difference between the results obtained by the two methods is minimal and the eigenvalue criteria are sufficient for each spacecraft data. Using the determined time intervals, the calculations of the ratio between the upstream and downstream total magnetic fields, ion plasma densities, and ion plasma temperatures as well as the bulk speed, and shock $\theta_{Bn}$ angle are made. These parameters, the MVA normal, and the magnetic CP normals are shown in Table~\ref{tab:core1}. The shock parameters fulfil the necessary shock criteria \citep{lumme17:_datab_helios_shock_waves}. The results of the additional parameter calculations are shown in Table~\ref{tab:core2}.

\begin{figure*}
\centering
\includegraphics[width=0.99\linewidth]{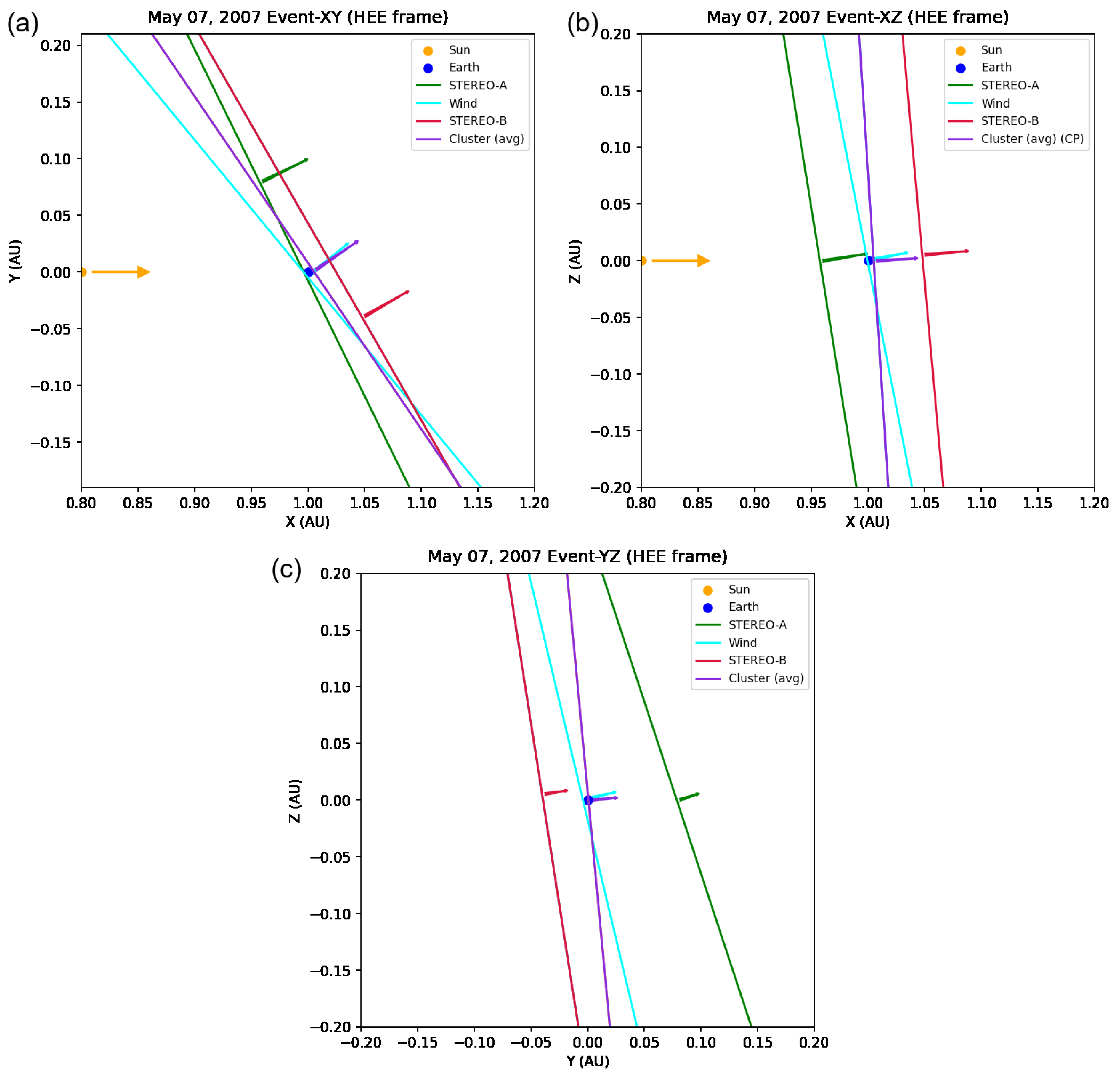}
\caption{2D (a) XY, (b) XZ, and (c) YZ sketches of the propagation of the IP shock through spacecraft -- Wind (light blue), STEREO$-$A (green) and B (red), and average position of four Cluster satellites (purple) on May 7, 2007. The orange arrow represents the Sun-Earth line direction. The arrows on spacecraft positions indicate the normal vector direction calculated from the co-planarity method and the lines perpendicular to the normal vectors indicate the shock surface orientations. The sizes of the lines are arbitrary. The shock orientations for the four Cluster satellites are averaged.}
\label{fig:05072D}
\end{figure*}

\begin{figure*}
\centering
\includegraphics[width=0.99\linewidth]{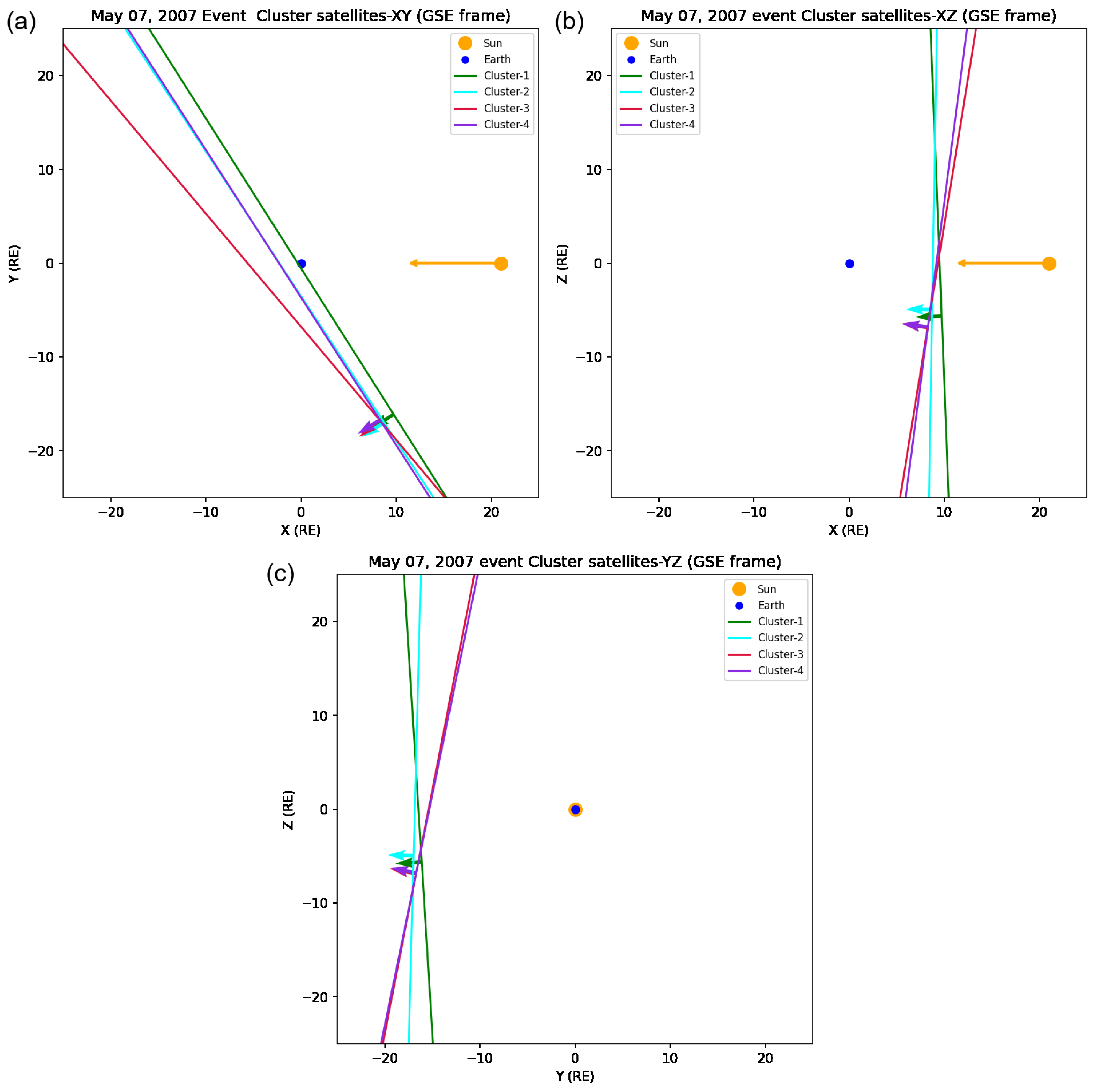}
\caption{2D sketches in (a) XY, (b) XY, and (c) YZ planes of the propagation of the IP shock through four Cluster satellites – SC1 (blue), SC2 (light blue), SC3 (purple), and SC4 (red) on May 7, 2007. The orange arrow represents the Sun-Earth line direction. The arrows on spacecraft positions indicate the normal vector direction calculated from co-planarity and the lines perpendicular to the normal vectors indicate the shock surface orientations. The sizes of the lines are arbitrary. \label{fig:clustertogether}}
\end{figure*}

Using the results, the 2D sketches of the IP shocks that were detected by the spacecraft are shown in Figure~\ref{fig:05072D}. In these 2D sketches, the shock propagation and normal vector orientation are shown in a temporal development manner. Since four Cluster satellites are relatively close to one another, their averaged position as well as normal vectors are shown in the general 2D sketches (also in 3D sketches, see later). For explicitly showing normal vector directions and positions of all Cluster satellites, it is suitable to change their coordinates and normal vectors into a GSE coordinate system with the positions in the Earth radii ($R_{E}$) unit, see Figure~\ref{fig:clustertogether}.

\begin{figure*}
\centering
\includegraphics[width=0.99\linewidth]{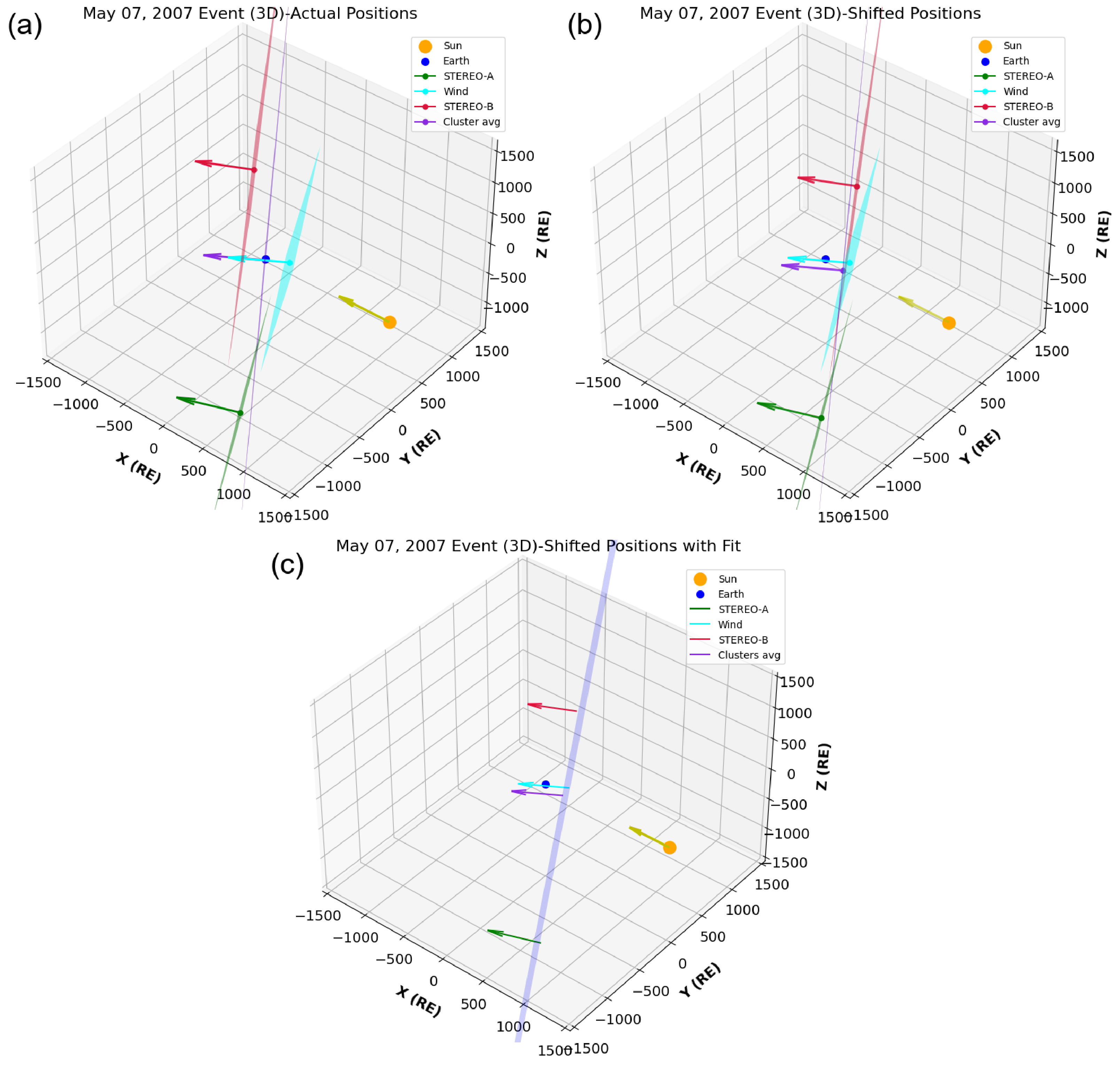}
\caption{A 3D sketch of the propagation of the IP shock through spacecraft-- Wind, STEREO$-$A, and STEREO$-$B, and average position of four Cluster satellites on May 7, 2007. (a) Actual shock detected positions. (b) The positions of STEREO$-$A, STEREO$-$B, and the average positions of four Cluster spacecraft are shifted back in time to the shock detection time of the Wind spacecraft. (c) The time-shifted positions are fitted with a plane. The arrows indicate the normal vector direction and the planes perpendicular to the normal vectors indicate the shock surface orientations. The sizes of the planes are arbitrary. \label{fig:05073D}}
\end{figure*}

The 3D sketches of the shock normals are shown in Figure~\ref{fig:05073D}. In order to assess the shape of the shock, we shifted the shock normals hence, as they had been detected contemporary with the shock observation time in Wind. For this, we used the average solar wind speeds and the time differences between the observations in Wind and the three other missions. The shifted shock normals are shown in Figure~\ref{fig:05073D}b. It turns out that the starting points of all of the normal vectors lay approximately in a plane surface, meaning that the underlying shock exhibits a planar structure in the spatial scale of the spacecraft distances. The plane fitted to the time-shifted normals is shown in Figure~\ref{fig:05073D}c.

\subsection{Event April 23, 2007} \label{sec:event20070423}

\begin{figure*}
\centering
\includegraphics[width=0.99\textwidth]{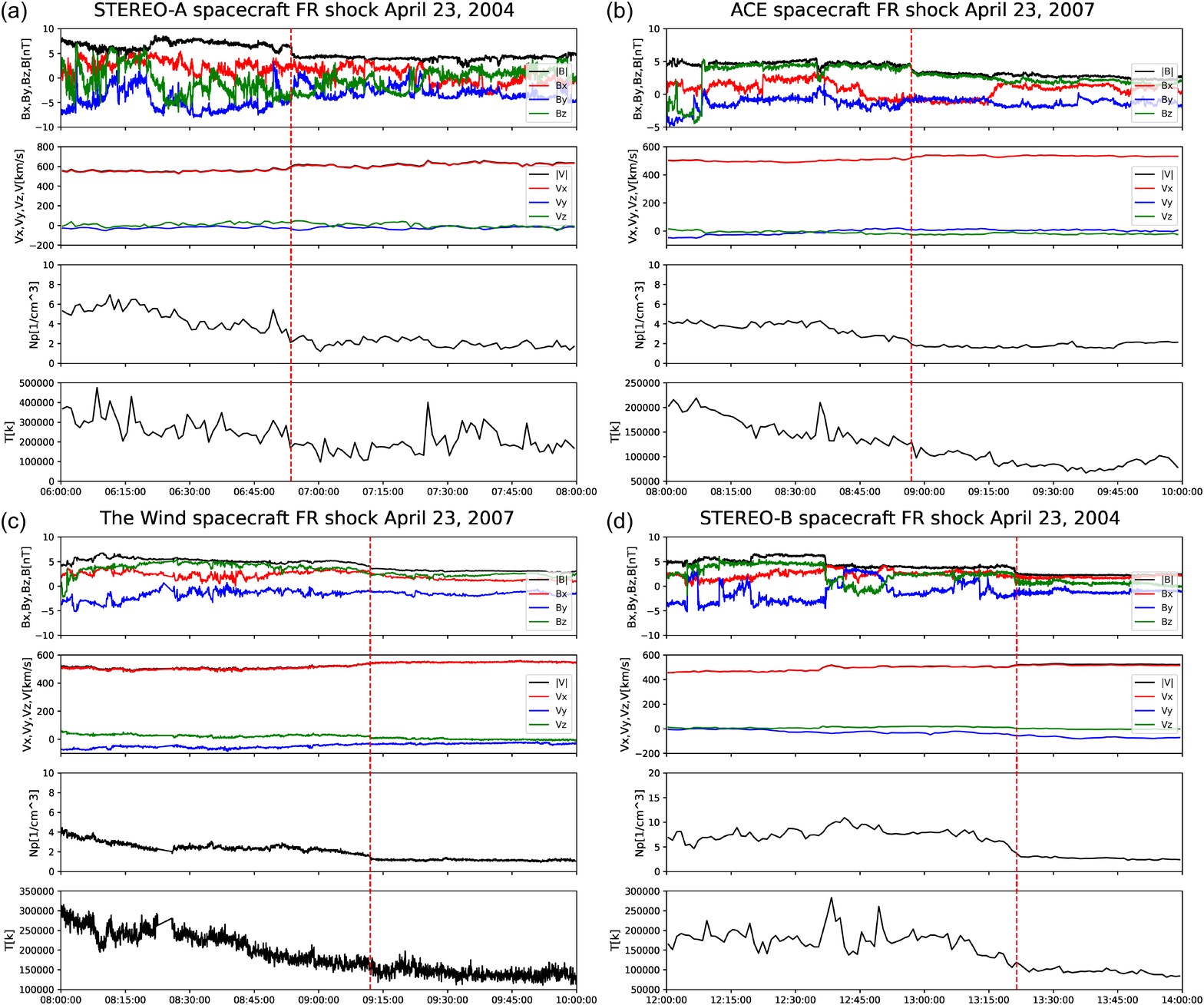}
\caption{(a) The plot of the shock detected by the STEREO$-$A spacecraft on April 23, 2007, at 06:53:35 (UTC). (b) The plot of the shock detected by the ACE spacecraft on April 23, 2007, at 08:57:00 (UTC). (c) The plot of the shock detected by the Wind spacecraft on April 23, 2007, at 09:12:00 (UTC). (d) The plot of the shock detected by the STEREO$-$B spacecraft on April 23, 2007, around 13:21:30 (UTC). FR stands for the fast reverse shock, which means the shock is moving toward its driver. The panels show from top to bottom, the magnetic field magnitude as well as its components, the total velocity and its components, density, and temperature. The dashed red line represents the exact shock time. The duration of the plot is two hours. \label{fig:ip20070423}}
\end{figure*}

In this specific event, the STEREO$-$A spacecraft detected a shock first at 06:53:35 (UTC), then the ACE spacecraft detected the shock at 08:57:00 (UTC), and the Wind spacecraft detected the shock at 09:12:00 (UTC). The magnetic field and the plasma data are shown in Figure~\ref{fig:ip20070423} for the STEREO$-$A (a), ACE (b), Wind (c), and STEREO$-$B (d).

STEREO$-$A and B have identical instruments, therefore, STEREO$-$B must have detected the shock on that day even though there is no detected IP shock for STEREO$-$B in the shock lists database mentioned in Section~\ref{sec:ipshock}. So, by using the average solar wind speed, 400\,km/s, we concluded that the shock detection time for STEREO$-$B should be around 13:00 to 15:00. Considering this, a possible shock signature from STEREO$-$B is found around 13:21:30 even though it is a faint signature. The magnetic field and the plasma plot are shown in Figure~\ref{fig:ip20070423}d.

\begin{figure*}
\centering
\includegraphics[width=0.99\textwidth]{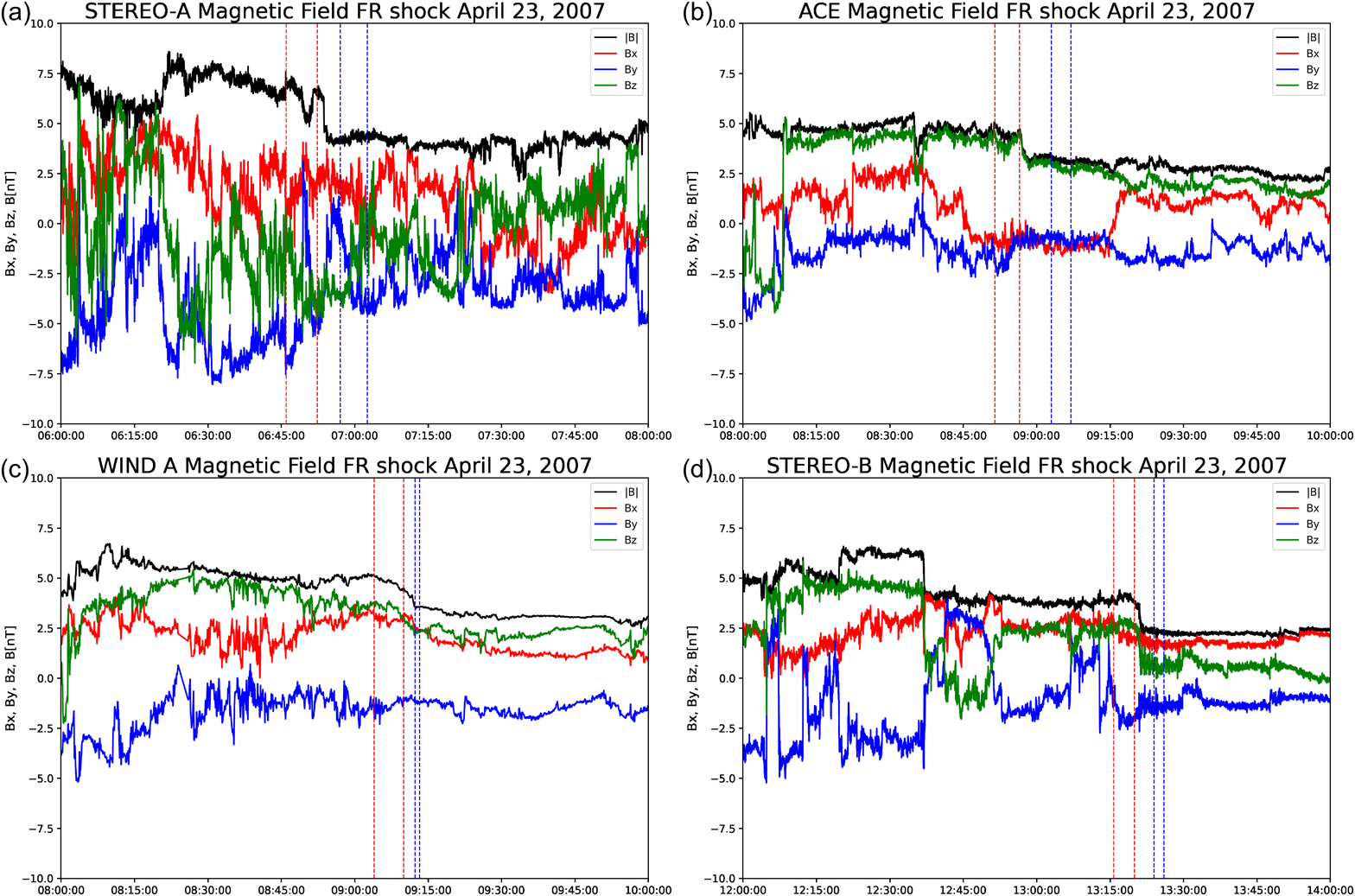}
\caption{(a) The STEREO$-$A magnetic field measurements on April 23, 2007. The downstream $\Delta t_{down}$ is between (06:46:00 - 06:52:23), and the upstream $\Delta t_{up}$ is between (06:57:03 - 07:02:34). (b) ACE magnetic field measurements on April 23, 2007. The downstream $\Delta t_{down}$ is between (08:51:30 - 08:56:30), and the upstream $\Delta t_{up}$ is between (09:03:00 - 09:07:00). (c) The Wind magnetic field measurements on April 23, 2007. The downstream $\Delta t_{down}$ is between (09:04:00 - 09:10:00), and the upstream $\Delta t_{up}$ is between (09:12:20 - 09:13:15). (d) STEREO$-$B magnetic field measurements on April 23, 2007. The downstream $\Delta t_{down}$ is between (13:15:45 - 13:20:00), and the upstream $\Delta t_{up}$ is between (13:24:00 - 13:26:00). FR stands for the fast reverse shock, which means the shock is moving toward its driver. The chosen intervals of the upstream and downstream magnetic fields were used for minimum variance analysis and co-planarity methods. The red dashed lines each represent the downstream starting time and ending time and the upstream starting time and ending time respectively. \label{fig:b20070423}}
\end{figure*}

We use all magnetic field observations of the spacecraft about the event and apply the MVA and the CP methods to them. In this event, the order between upstream and downstream is swapped because it is a fast-reverse (FR) shock event, which means the shock propagates toward the Sun against the flow but is carried outward from the Sun by the bulk solar wind flow \cite{kilpua15:_proper_earth}. The upstream and downstream time intervals that give the best agreement between the results of the MVA and CP shock normal determination for all spacecraft are shown in Table~\ref{tab:timeintervalsapril}. Figure~\ref{fig:b20070423} shows the corresponding magnetic field observations for STEREO$-$A (a), ACE (b), Wind (c) and STEREO$-$B (d) missions. Table~\ref{tab:timeintervalsapril} also shows the ratio between the intermediate eigenvalue $\lambda_{2}$ and the smallest eigenvalue $\lambda_{3}$ as well as the angle between the MVA and CP normal vector determinations of the shock in the upstream and downstream time intervals for each spacecraft.

Similarly to the Event May 7, the upstream and downstream time intervals are defined correctly as well, considering the angle difference between the two methods is minimal and the eigenvalue criteria are fulfilled for each spacecraft data. Also, using the determined time intervals, in Table~\ref{tab:coreapril1} we summarize, for each mission, the ratio between the upstream and downstream total magnetic fields, densities and temperatures, and the differences between the upstream and downstream solar wind bulk speeds. From the obtained upstream/downstream ratios and bulk speed differences we conclude that the criteria for detecting a shock event in the considered period is fulfilled \citep{lumme17:_datab_helios_shock_waves} for each spacecraft. Even for STEREO$-$B, for which shock was not indicated in the shock catalogue (see above), the numbers unambiguously prove the existence of a shock event during the studied interval. However, the almost quasi-parallel shock detected by the STEREO$-$B appears to become quasi-parallel a few hours after the shock detection times of STEREO$-$A, ACE, and the Wind based on the shock $\theta_{Bn}$ angle, see Table~\ref{tab:coreapril1}, where the calculated results of additional parameters are shown.

\begin{figure*}
\centering
\includegraphics[width=0.99\linewidth]{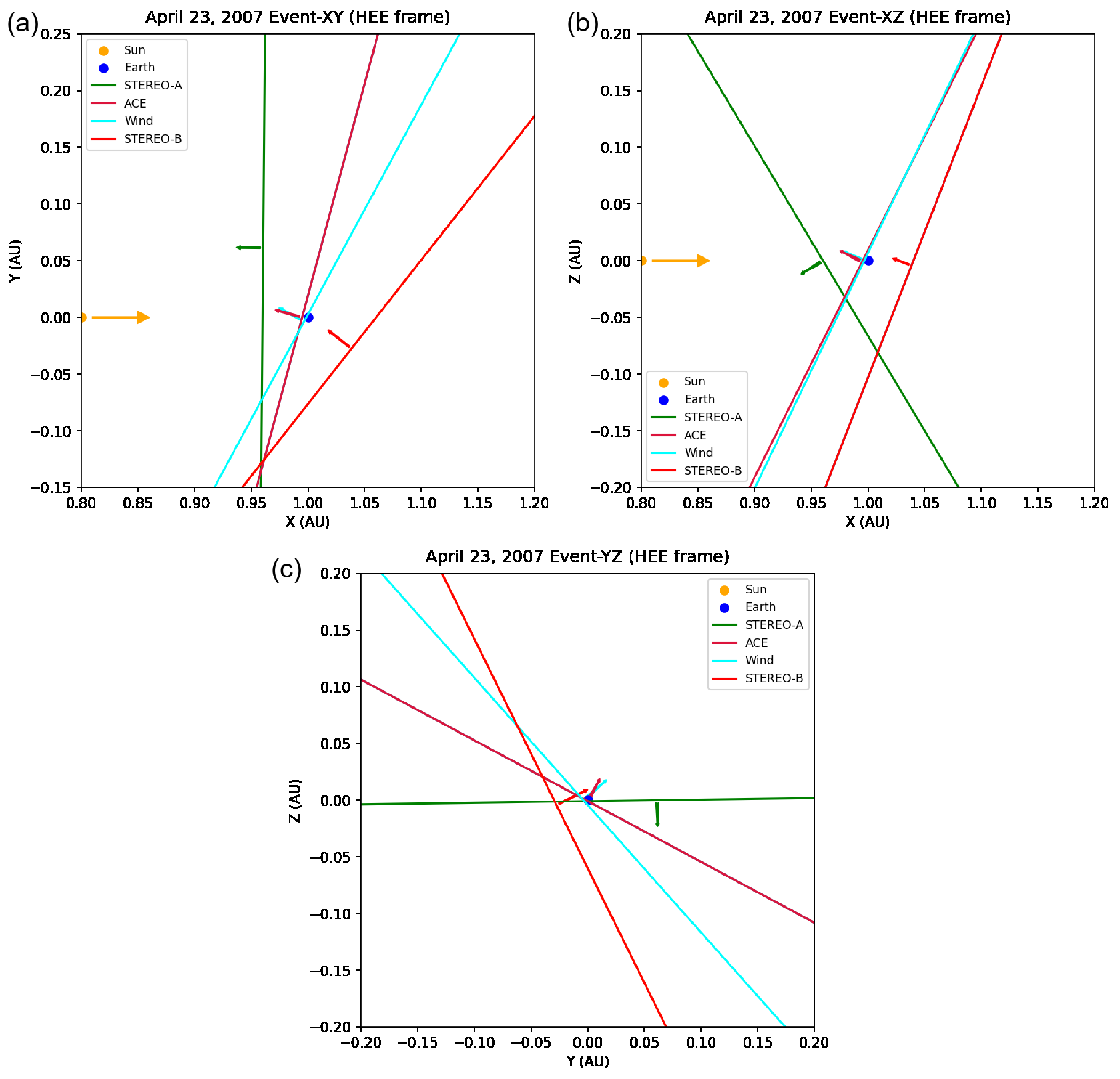}
\caption{2D (a) XY, (b) XZ, and (c) YZ sketches of the propagation of the IP shock through spacecraft-- STEREO$-$A, ACE, Wind, and STEREO$-$B. The orange arrow represents the Sun-Earth line direction. The arrows on spacecraft positions indicate the normal vector direction calculated from the co-planarity method and the lines perpendicular to the normal vectors indicate the shock surface orientations. The sizes of the lines are arbitrary. \label{fig:04232D}}
\end{figure*}

The sketches of the IP shock geometries that were detected by the spacecraft are shown in Figure~\ref{fig:04232D}. In these 2D sketches, the shock propagation and normal vector orientation are shown in a temporal development manner.

The 3D sketch of the IP shock geometries is shown in Figure~\ref{fig:04233D}. The 3D sketch, the STEREO$-$A, and B positions are time-shifted to the position of STEREO$-$A to see the overall shape of this IP shock. In order to assess the shape of the shock, we shifted the shock normals hence, as they had been detected contemporary with the shock observation time in STEREO$-$A. For this, we used the average solar wind speeds and the time differences between the observations in STEREO$-$A and in the two other spacecraft. The shifted shock normals are shown in the top right panel of Figure~\ref{fig:04233D}. The top view of the time-shifted normals is shown in the bottom part of Figure~\ref{fig:04233D}.

\section{Discussion}
\label{sec:disc}

The IP shock parameters did not change significantly from STEREO$-$A to B in either case (Table~\ref{tab:core1}, \ref{tab:core2},~\ref{tab:coreapril1}, \ref{tab:coreapril2}). The IP shocks did not evolve on the scale of 40\,million\,km, that is the distance of the STEREO spacecraft.

\subsection{Event May 7, 2007} \label{sec:disc20070507}

\begin{figure*}[!t]
\centering
\includegraphics[width=0.99\textwidth]{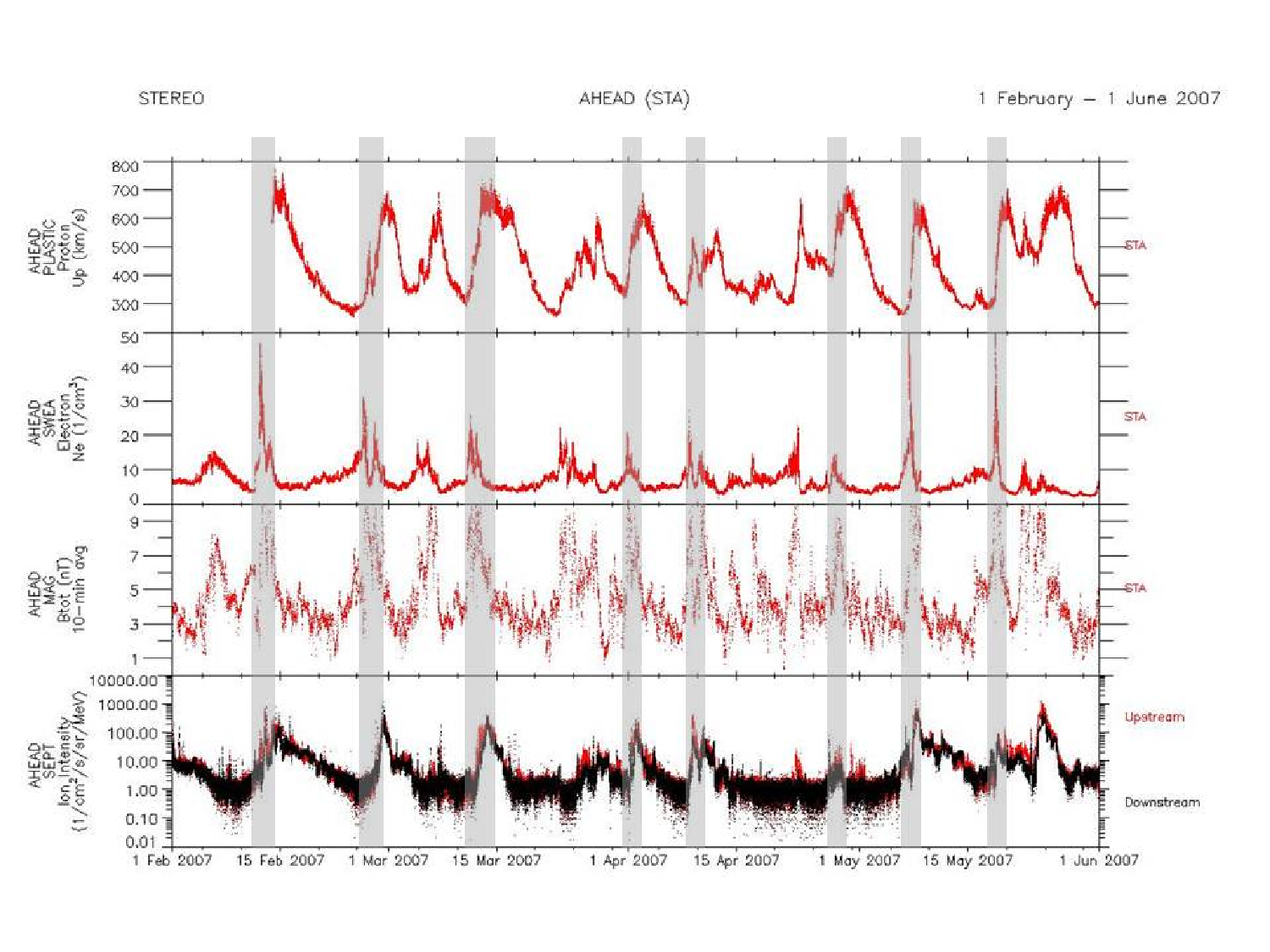}
\caption{STEREO$-$A PLASTIC bulk velocity, the SWEA electron density, MAG magnetic field magnitude, and the SEPT ion intensity from 110\,keV to 2200\,keV. From February 1, 2007, to June 1, 2007, periods featuring well-established Co-rotating Interaction Regions (CIRs) with noticeable boosts in ion energy are highlighted with grey shading \cite[Figure~2]{opitz14:_solar_stereo}. \label{fig:cirpaper}}
\end{figure*}

\begin{figure*}[!t]
\centering
\includegraphics[width=0.99\textwidth]{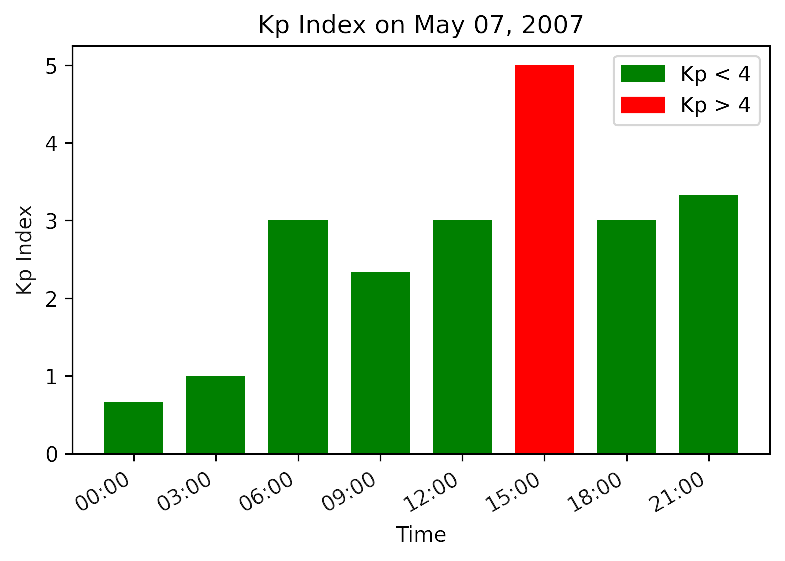}
\caption{$K_{p}$-index on May 07, 2007. The green bars indicate moderate geomagnetic activity, while the red bar indicates a G1-minor geomagnetic storm. \label{fig:kp0507}}
\end{figure*}

The IP shock fitted plane is tilted $56.42^{\circ}$ from the Sun-Earth line according to Figure~\ref{fig:05073D}c. The tilt appears to be almost the same as expected for the Parker spiral impacting the Earth from the dawn side. Hence, it is the reason why the Wind detected the shock first even though STEREO$-$A's position is relatively closer to the Sun's direction. This period, 2007, was during the solar minimum phase, and stream interaction region (SIR) or corotating interaction region (CIR) were dominant \citep{opitz14:_solar_stereo}, and this further proves the result, see Figure~\ref{fig:cirpaper}. Furthermore, to see a correlation between fast-forward shock and geomagnetic activity, we investigated the $K_{p}$-index \citep[\url{https://www.swpc.noaa.gov/noaa-scales-explanation}]{bartels49:_index_ks_planet_kp_iatme_bull,rostoker72:_geomag,matzka21:_geomag_kp_index_deriv_indic_geomag_activ} on May 07, 2007, as shown in Figure~\ref{fig:kp0507}. It appears that this forward shock-leading event disturbed the magnetosphere, causing a G1-minor geomagnetic storm with the $K_{p}$- index peaking around 15:00 (UTC). The geomagnetic substorm happened about 6 hours after the detection of the shock by the STEREO$-$B spacecraft.

\subsection{Event April 23, 2007} \label{sec:disc20070423}

\begin{figure*}
\centering
\includegraphics[width=.79\linewidth]{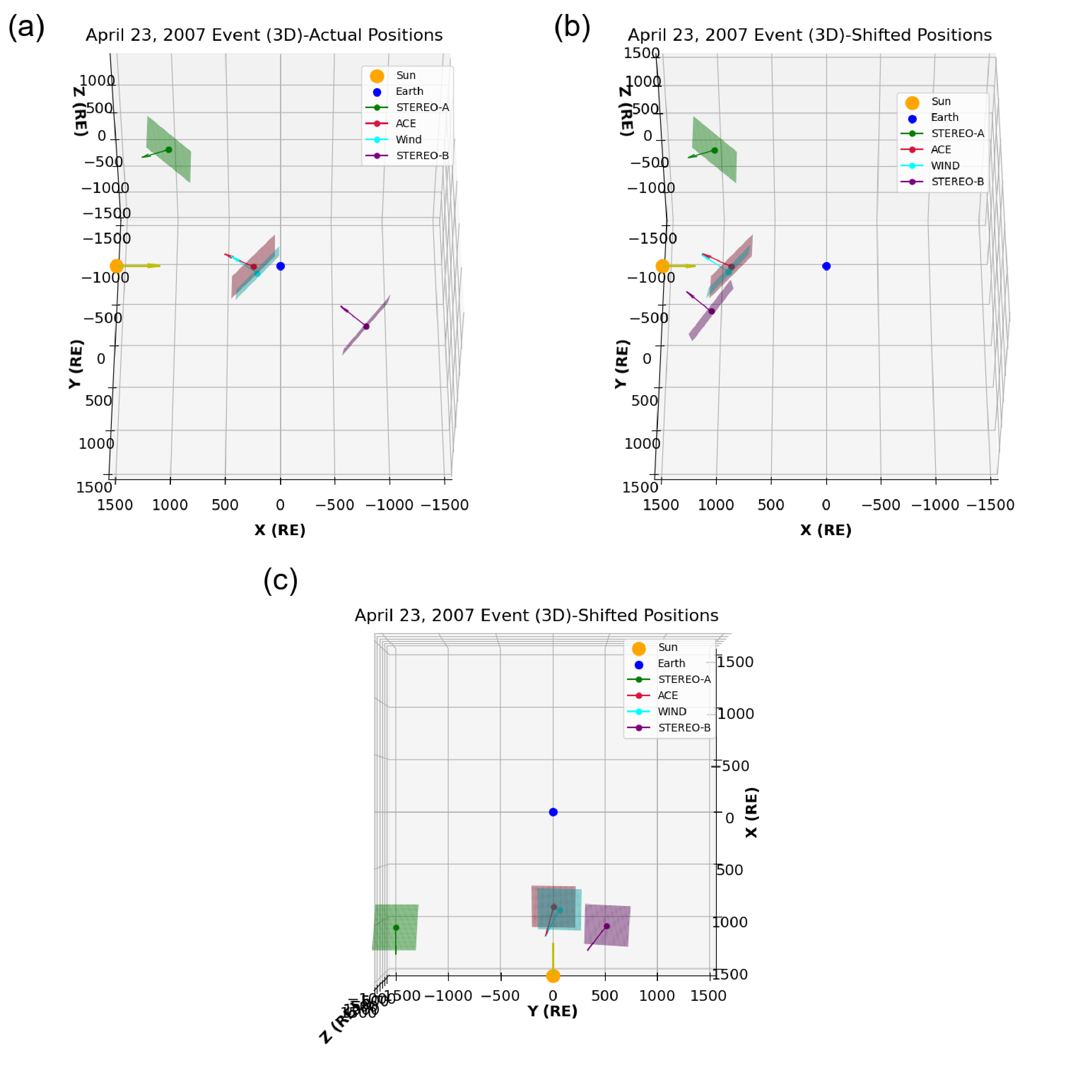}
\caption{(a) A 3D sketch of the propagation of the IP shock through four spacecraft-- STEREO$-$A, Wind, ACE, and STEREO$-$B. (b) The positions of Wind, ACE, and STEREO$-$B are shifted back in time to the shock detection time of STEREO$-$A. (c) The shifted positions are shown from the top view. The arrows indicate the normal vector direction and the planes perpendicular to the normal vectors indicate the shock surface orientations. The sizes of the planes are arbitrary. \label{fig:04233D}}
\end{figure*}

In Figure~\ref{fig:04233D} it seems that the shape of the shock is not uniform and is twisted from the STEREO$-$A spacecraft to the other three spacecraft about the Z-axis along the transverse direction (Y-axis) to the Sun-Earth line, see Figure~\ref{fig:04233D}b. As seen from Figure~\ref{fig:04233D}c, the STEREO$-$A alone is on the dawn (left) side of the Sun-Earth line while the Wind, the ACE, and the STEREO$-$B are on the dusk (right). In this XY-plane view (Figure~\ref{fig:04233D}), the shock normal vectors appear to be changing or slightly rotating their direction from the STEREO$-$A to the Wind, the ACE, and the STEREO$-$B about the Z-axis, and the shock is changing from the quasi-perpendicular to quasi-parallel based on the shock $\theta_{Bn}$ angle. This IP shock is a fast reverse shock, meaning it travels toward the Sun in a frame fixed to the solar wind even though the shock propagates away from the Sun with the solar wind. The shock detection time difference between the STEREO$-$A and the Wind/ACE is about two hours while between STEREO$-$A and B is almost six hours. Yet the shock orientation is significantly changed from the STEREO$-$A to the other three spacecraft indicating the change is spatial, not temporal. So, due to this nature, the IP shock could be a local ripple. The ripples on the shock surface are known to be caused by ICME (interplanetary coronal mass ejections) shock drivers as they do not propagate into homogeneous interplanetary medium \citep{acuna08:_stereo_impac_magnet_field_exper}. However, for this particular event the source is suggested to be a stream interaction region (SIR) according to the catalogue given in \url{https://stereo-ssc.nascom.nasa.gov/pub/ins_data/impact/level3/STEREO_Level3_Shock.pdf}.

\begin{figure*}[!t]
\centering
\includegraphics[width=0.99\textwidth]{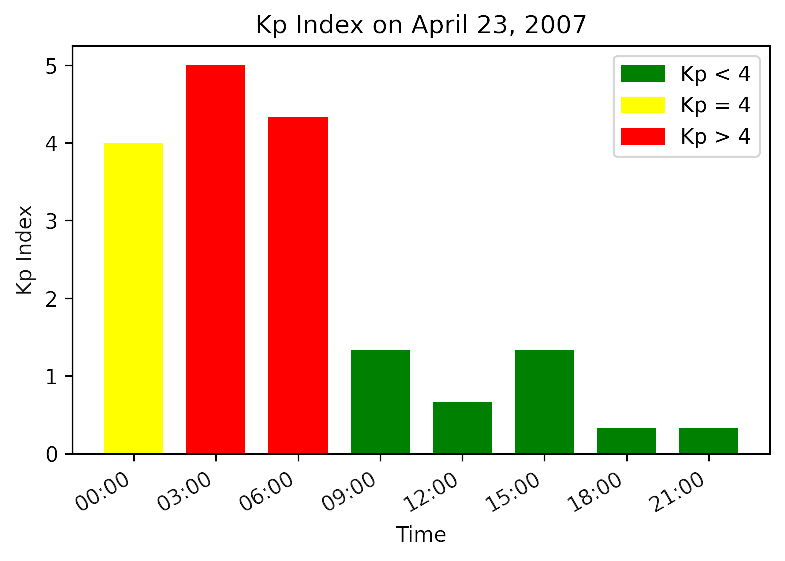}
\caption{The $K_{p}$-index on April 23, 2007. The green bars indicate moderate geomagnetic activity, while the yellow bar denotes intensifying geomagnetic activity with the red bars symbolize a geomagnetic storm. \label{fig:kp0423}}
\end{figure*}

We investigated the $K_{p}$-index of this event as seen in Figure~\ref{fig:kp0423}. Similar to the fast-forward shock event of May 07, 2007, a G1-minor geomagnetic storm occurred on this day (see space weather scales of National Oceanic and Atmospheric Administration (NOAA) \url{https://www.swpc.noaa.gov/noaa-scales-explanation}), but the beginning of the geomagnetic storm happened three hours before the shock detection time of the STEREO$-$A while the ending of the storm was partially at the same time as the STEREO$-$A shock detection. As stated before, SIR/CIRs form when the fast-moving solar wind catches the slow-moving solar wind \cite{richardson18:_solar}. This sometimes forms a pair of shocks, one leading as a fast-forward shock while a rarefied shock trails the solar wind as a fast-reverse shock but oftentimes as just sole fast-forward or fast reverse shocks \citep{jian19:_solar_terres_relat_obser_stereo}. Nevertheless, since it always involves the fast-moving solar wind, it makes total sense why the G1-minor geomagnetic storm with a $K_{p}=5$ index happened before the detection of the shock because it seems the fast-moving solar wind caused the minor storm before the fast-reverse shock finally arrived at the spacecraft positions.

The normal orientation inconsistency can also be explained by the argument that this shock is a fast-reverse shock because SIRs have characteristic tilts such that forward waves direct towards the solar equatorial plane while the reverse waves tend to move in the direction of the solar poles \citep{kilpua15:_proper_earth}. Therefore, this tilting nature may explain the XZ component's tilted orientation between the STEREO$-$A and the other three spacecraft -- the Wind, the ACE, and the STEREO$-$B.

\section{Summary and conclusions} \label{sec:sum}

In this paper, two conjugated multi-spacecraft studies of interplanetary (IP) shock propagation are presented. It aims to determine how IP shocks develop and evolve in spatial and temporal propagation. The spatial separation of observations is around 40\,million\,km. In this scale, we do not see any significant development of the IP shocks. The shock parameters are similar within their estimated error at each location. No sign of wave dispersion was observed.

The Event May 7, 2007, is a fast-forward (FF) shock event, appears planar and tilted $56.42^{\circ}$ to the Sun-Earth line. The tilt of this planar shock surface is almost identical to the usual propagation of the Parker spiral impacting the Earth. Therefore, the origin of the shock is corotating interaction region (CIR), which agree with the detected CIR on that day. There is no sign of temporal change in this scale. In the spatial range of 40\,million km, no change was observable. The G1-minor geomagnetic storm happened after this shock was detected, indicating this shock event caused the geomagnetic storm on that day.

The Event of April 23, 2007 is a fast reverse (FR) shock event and, it appears that the shape of the shock is not uniform and twisted from the STEREO$-$A spacecraft to the other spacecraft along the transverse direction to the Sun-Earth line. The shock normal vectors also were observed to be changing along the Y-axis. A G1-minor geomagnetic storm occurred on this day, similar to the fast-forward shock event of May 7, 2007. Intriguingly, the onset of this storm took place three hours before the shock detection time recorded by STEREO$-$A. The termination of the storm partially coincided with the STEREO$-$A detection time. The source of this shock is SIR/CIR. The G1-minor geomagnetic storm, with a $K_{p}$-5 index, began before the shock detection. It seems that the fast-moving solar wind instigated the minor storm before the arrival of the fast-reverse shock at the spacecraft's position. The orientation irregularity can be potentially accounted for by the characteristic tilts of SIRs.

No multipoint observation was made before using so many and so different probes. The 3D spatial curve and the origin of the IP shocks are determined. These macroscopic features were visible only in heliospheric magnetohydrodynamic simulations before. However, the known IP shock catalogues of the STEREO, the Wind, the ACE and the Cluster missions provided older observations. A natural continuation of this study is using PSP, the SolO, and the BepiColombo \citep{milillo10:_bepic} measurements for joint observations. The larger spatial separation of the IP shock observations might allow us to observe the temporal development and dispersion of these shocks.

%%% End of body of article

%%%%%%%%%%%%%%%%%%%%%%%%%%%%%%%%%%%%%%%%%%%%%%%
%
% DATA SECTION and ACKNOWLEDGMENTS
%
%%%%%%%%%%%%%%%%%%%%%%%%%%%%%%%%%%%%%%%%%%%%%%%

\begin{acknowledgments}
\textbf{Acknowledgments} This paper uses data from the Heliospheric Shock Database, generated and maintained at the University of Helsinki (\url{https://www.ipshocks.fi}). We would also like to acknowledge NASA CDAWeb, the STEREO$-$A and B IMPACT, PLASTIC, the Wind MFI, 3DP, ACE MAG, SWEPAM, and the Cluster FGM, CIS, EFW teams for providing data, and TREPS, designed and developed by the French Plasma Physics Data Centre (CDPP), for coordinate transformations for this study. This work was partially financed by the National Research, Development, and Innovation Office (NKFIH) FK128548 grant. ML was supported by the Stipendium Hungaricum Scholarship. The authors thank Oleksiy Agapitov, G{\'e}za Erd{\H{o}}s and Zolt{\'a}n N{\'e}meth for the useful discussions.
\end{acknowledgments}
%
%%%%%%%%%%%%%%%%%%%%%%%%%%%%%%%%%%%%%%%%%%%%%%%%
%% REFERENCES and BIBLIOGRAPHY
%%
%%
%%%%%%%%%%%%%%%%%%%%%%%%%%%%%%%%%%%%%%%%%%%%%%%%

%\pagebreak

\bibliography{aphys-2024-ipshocks-tx-arxiv}{}
\bibliographystyle{aasjournal}

\vfill

\pagebreak

\begin{longrotatetable}
\begin{deluxetable*}{ccccccc}
\tablecaption{The parameters and main results of the minimum variance (MVA) and magnetic co-planarity (CP) analyses of magnetic field records of the considered missions recorded about the May 7, 2007 shock event. Here, $\Delta t_{up}$ denotes the defined upstream and $\Delta t_{down}$ denotes the defined downstream time intervals with MVA and CP analysis methods. $\mathbf{n}_{MVA} $ and $\mathbf{n}_{CP}$ are the normal vectors given by MVA and CP methods, respectively. $\lambda_{2}/\lambda_{3}$ indicates the ratio between the intermediate eigenvalue $\lambda_{2}$ and the smallest eigenvalue $\lambda_{3}$, and $\Delta\theta_{MVA-CP}$ is the angle between the MVA and CP normals on May 7, 2007. \label{tab:timeintervalsmay}}

\tablehead{
  \colhead{Spacecraft} & \colhead{$\Delta t_{up}$} & \colhead{$\Delta t_{down}$} & \colhead{$\mathbf{n}_{MVA}$} & \colhead{$\mathbf{n}_{CP}$} & \colhead{$\lambda_{2}/\lambda_{3}$} & \colhead{$\Delta\theta_{MVA-CP}$}\\
 & \colhead{$\left[t_{up}^{start}-t_{up}^{end}\right]$} & \colhead{$\left[t_{down}^{start}-t_{down}^{end}\right]$} & & \\
 & \colhead{[hh:mm:ss-hh:mm:ss]} & \colhead{[hh:mm:ss-hh:mm:ss]} & & & & \colhead{$\left[{}^{\circ}\right]$}
 }
  \startdata
Wind & 06:59:00 - 07:01:47 & 07:03:50 - 07:05:50 & [-0.76, -0.63, -0.14] & [-0.76, -0.63, -0.15] & 14.82 & 0.38\\
STEREO$-$A & 08:08:00 - 08:10:05& 08:12:30 - 08:13:50 & [0.89, 0.44, 0.14] & [0.89, 0.44, 0.144] & 3.16 & 0.37\\
STEREO$-$B & 09:37:00 - 09:40:35& 09:42:57 - 09:43:38 & [-0.85 -0.51 -0.09] & [-0.86 -0.50 -0.08] & 9.51 & 1.10 \\
Cluster-1 & 08:26:10 - 08:27:00& 08:28:00 - 08:29:00& [0.85 0.53 -0.03] & [-0.85 -0.53 -0.03] & 167.90 & 0.21 \\
Cluster-3 & 08:26:30 - 08:27:10& 08:29:00 - 08:29:47 & [0.75 0.64 0.15] & [-0.76 -0.63 -0.12] & 22.43 & 1.92 \\
Cluster-2 & 08:26:56 - 08:27:40& 08:28:32 - 08:29:37 & [0.84 0.54 0.01] & [-0.84 -0.54 -0.01] & 91.56 & 0.10\\
Cluster-4 & 08:26:25 - 08:27:10& 08:28:45 - 08:29:40& [0.86 0.51 -0.08] & [0.84 0.53 0.11] & 238.63 & 2.49 \\
\enddata
\end{deluxetable*}
\end{longrotatetable}

\pagebreak

\begin{longrotatetable}
\begin{deluxetable*}{cccccc}
\tablecaption{Resulting core parameters of studying the data of the Wind, STEREO$-$A, B, and Cluster spacecraft on May 7, 2007. $B_{d}/B_{u}$, $N_{d}/N_{u}$, and $T_{d}/T_{u}$ are the ratio of the upstream and downstream magnetic field, density, and temperature, respectively. $\Delta V$ is the velocity difference between the two sides of the shock. Furthermore, $\theta_{Bn}$ is the shock $\theta$ angle. Due to the CIS experiment is not operational for Cluster SC2, and the CIS HIA sub$-$instrument being switched off for Cluster SC4 spacecraft, the plasma parameters are not available for these satellites. The Cluster CIS HIA instruments SC1 and SC3 provide parallel ($T_{\parallel}$) and perpendicular ($T_{\perp}$) temperatures, respectively. \label{tab:core1}}
\tablehead{
\colhead{Spacecraft} & \colhead{$B_{d}/B_{u}$} & \colhead{$N_{d}/N_{u}$} & \colhead{$T_{d}/T_{u}$} & \colhead{$\Delta V$} & \colhead{$\theta_{Bn}$} \\
   & & & & \colhead{[km/s]} & \colhead{$\left[{}^{\circ}\right]$} \\
}
\startdata
Wind & $1.86\pm 0.10$ & $1.70\pm 0.06$ & $1.95\pm 0.12$ & $26.8\pm 1.7$ & 70.80 \\
STEREO$-$A & $1.89\pm 0.02$ & $2.18\pm 0.90$ & $3.21\pm 2.4$ & $36.00\pm 6.00$ & 81.83 \\
STEREO$-$B & $1.90\pm 0.07$ & $1.90\pm 0.6$ & $2.90\pm 6.00$ & $40.25\pm 11.00$ & 59.45 \\
 Cluster 1 & $1.53\pm 0.04$ & $1.63\pm 0.09$ & $1.10_{\parallel}\pm 1.30_{\parallel}$ & $31.20\pm 3.10$ & 86.62 \\
&&&$1.40_{\perp}\pm 0.32_{\perp}$ & &\\
Cluster 3 & $1.57\pm 0.05$ & $1.69\pm 0.08$ & $1.20_{\parallel}\pm 1.50_{\parallel}$ & $30.50\pm 2.80$ & 89.74 \\
&&&$1.40_{\perp}\pm 0.40_{\perp}$&&\\
Cluster 2 & $1.53\pm 0.03$ & $1.42\pm 0.18$ & N/A & N/A & 88.19 \\
Cluster 4 & $1.57\pm 0.05$ & $1.49\pm 0.19$ & N/A & N/A & 86.55 \\
\enddata
\end{deluxetable*}
\end{longrotatetable}

\pagebreak

\begin{longrotatetable}
\begin{deluxetable*}{ccccccccc}
\tablecaption{Resulting additional parameters of studying the data of the three spacecraft on May 7, 2007. $V_{sh}$ is the shock speed, $C^{up}_s$ is the upstream sound speed, $V^{up}_A$ is the upstream Alfvén speed, and $C^{up}_{ms}$ is the upstream magnetosonic speed. Plasma $\beta_{up}$ based on upstrem parameters, Alfvén-Mach ($M_{A}$), and Magnetosonic-Mach ($M_{ms}$) numbers are also shown.\label{tab:core2}}
\tablehead{
  \colhead{Spacecraft} & \colhead{$V_{sh}$ [km/s]} & \colhead{$C^{up}_s$ [km/s]} & \colhead{$V^{up}_A$ [km/s]} & \colhead{$C^{up}_{ms}$ [km/s]} & \colhead{Plasma $\beta_{up}$} & \colhead{$M_{A}$} & \colhead{$M_{ms}$} \\
  }
\startdata
Wind & 317.33 & $48.84\pm 2.73$ & $29.81\pm 10.91$ & $57.22\pm 6.14$ & $3.22\pm 2.38$ & $2.83\pm 1.20$ & $1.47\pm 0.35$ \\
STEREO$-$A & 352.41 & $46.27\pm 3.30$ & $34.89\pm 15.00$ & $57.95\pm 9.41$ & $2.11\pm 1.80$ & $2.19\pm 1.12$ & $1.32\pm 0.42$ \\
STEREO$-$B & 354.91 & $46.89\pm 5.85$ & $31.95\pm 7.33$ & $56.74\pm 6.35$ & $2.58\pm 1.34$ & $2.76\pm 1.07$ & $1.55\pm 0.52$ \\
Cluster 1 & 342.02 & $43.88\pm 3.09$ & $25.59\pm 7.70$ & $51.31\pm 3.99$ & $3.26\pm 1.89$ & $3.84\pm 1.40$ & $1.99\pm 0.46$ \\
Cluster 3 & 308.11 & $43.88\pm 2.63$ & $26.80\pm 7.64$ & $51.42\pm 3.98$ & $3.21\pm 1.83$ & $3.72\pm 1.35$ & $1.93\pm 0.46$ \\
\enddata
\end{deluxetable*}
\end{longrotatetable}

\pagebreak

\begin{longrotatetable}
\begin{deluxetable*}{ccccccc}
\tablecaption{The parameters and main results of the minimum variance (MVA) and co-planarity (CP) analyses of magnetic field records of the considered missions recorded about the April 23, 2007 shock event. Here, $\Delta t_{down}$ denotes the defined downstream and $\Delta t_{up}$ denotes the defined upstream time intervals with MVA and CP analysis methods. $\mathbf{n}_{MVA} $ and $\mathbf{n}_{CP}$ are the normal vectors given by MVA and CP methods, respectively. $\lambda_{3}/\lambda_{2}$ indicates the ratio between the intermediate eigenvalue $\lambda_{2}$ and the smallest eigenvalue $\lambda_{3}$, and $\Delta\theta_{MVA-CP}$ is the angle between the MVA and CP normals on April 23, 2007.\label{tab:timeintervalsapril}}
\tablehead{
  \colhead{Spacecraft} & \colhead{$\Delta t_{down}$} & \colhead{$\Delta t_{up}$} & \colhead{$\mathbf{n}_{MVA} $} & \colhead{$\mathbf{n}_{CP}$} & \colhead{$\lambda_{2}/\lambda_{3}$} & \colhead{$\Delta\theta_{MVA-CP}$} \\
    & \colhead{$\left[t_{down}^{start}-t_{down}^{end}\right]$} & \colhead{$\left[t_{up}^{start}-t_{up}^{end}\right]$} & & \\
& \colhead{[hh:mm:ss-hh:mm:ss]} & \colhead{[hh:mm:ss-hh:mm:ss]} & & & & \colhead{$\left[{}^{\circ}\right]$}\\
}
\startdata
STEREO$-$A & 06:46:00 - 06:52:23 & 06:57:03 - 07:02:34 & [-0.86, 0.02, -0.50] & [-0.86, 0.01, -0.51] & 3.29 & 1.07\\
ACE & 08:51:30 - 08:56:30 & 09:03:00 - 09:07:00 & [0.87, -0.19, -0.45] & [0.87, -0.23, -0.43] & 4.22 & 3.43\\
Wind & 09:04:00 - 09:10:00 & 09:12:20 - 09:13:15 & [-0.82, 0.43, 0.37] & [-0.81, 0.44, 0.39] & 3.38 & 1.46\\
STEREO$-$B & 13:15:45 - 13:20:00 & 13:24:00 - 13:26:00 & [-0.75, 0.59, 0.29] & [-0.73, 0.61, -0.30] & 8.18 & 1.67\\
\enddata
\end{deluxetable*}
\end{longrotatetable}

\pagebreak

\begin{table*}[ht]
 \centering
 \caption{Resulting core parameters of studying the data of the STEREO$-$A, B, ACE, and Wind spacecraft on April 23, 2007. $B_{d}/B_{u}$, $N_{d}/N_{u}$, and $T_{d}/T_{u}$ are the ratio of the upstream and downstream magnetic field, density, and temperature, respectively. $\Delta V$ is the velocity difference on the sides of the shock. Here $\theta_{Bn}$ is the shock $\theta$ angle. \label{tab:coreapril1}}
\begin{tabular}{ccccccccc}
\hline
 Spacecraft & $B_d/B_u$ & $N_d/N_u$ & $T_d/T_u$ & $\Delta V $ & $\theta_{Bn}$ \\
 & & & & [km/s] & $\left[{}^{\circ}\right]$ \\
 \hline
STEREO$-$A & $1.47\pm 0.16$ & $2.00\pm 0.70$ & $1.48\pm 0.50$ & $46.00\pm 13$ & 88.40 \\
ACE & $1.41\pm 0.08$ & $1.44\pm 0.16$ & $1.22\pm 0.09$ & $23.72\pm 5.57$ & 46.2 \\
Wind & $1.35\pm 0.06$ & $1.51\pm 0.11$ & $1.06\pm 0.12$ & $19.00\pm 6.00$ &61.85 \\
STEREO$-$B & $1.73\pm 0.09$ & $1.98\pm 0.14$ & $1.34\pm 0.15$ & $21.20\pm 3.40$ & 29.39 \\
\end{tabular}
\end{table*}

\pagebreak

\begin{longrotatetable}
\begin{deluxetable*}{ccccccccc}
\tablecaption{Resulting additional parameters of studying the data of the three spacecraft on April 23, 2007. $V_{sh}$ is the shock speed, $C^{up}_s$ is the upstream sound speed, $V^{up}_A$ is the upstream Alfvén-speed and $C^{up}_{ms}$ is the upstream magnetosonic speed. Plasma $\beta_{up}$, Alfvén$-$Mach number, and Magnetosonic Mach number are also shown.}\label{tab:coreapril2}
\tablehead{
  \colhead{Spacecraft} & \colhead{$V_{sh}$} & \colhead{$C^{up}_s$} & \colhead{$V^{up}_A$} & \colhead{$C^{up}_{ms}$} & \colhead{Plasma $\beta_{up}$} & \colhead{Alfvén-Mach} & \colhead{Magnetosonic-Mach} \\
& \colhead{[km/s]} & \colhead{[km/s]} & \colhead{[km/s]} & \colhead{[km/s]} & & &  \\  
}
\startdata
STEREO$-$A & 467.95 & $64.97\pm 7.74$ & $68.71\pm 29.20$ & $94.57\pm 21.85$ & $1.07\pm 0.94$ & $1.00\pm 0.67$ & $0.72\pm 0.41$ \\
ACE & 412.36 & $58.28\pm 4.94$ & $52.82\pm 15.90$ & $78.66\pm 11.26$ & $1.46\pm 0.91$ & $1.49\pm 0.56$ & $1.00\pm 0.27$ \\
Wind & 395.53 & $63.79\pm 5.15$ & $69.16\pm 21.71$ & $94.09\pm 16.34$ & $1.02\pm 0.66$ & $0.90\pm 0.42$ & $0.66\pm 0.26$ \\
STEREO$-$B & 370.31 & $57.31\pm 5.38$ & $29.04\pm 12.28$ & $64.25\pm 07.34$ & $4.67\pm 4.04$ & $1.69\pm 1.08$ & $0.76\pm 0.37$ \\
\enddata
\end{deluxetable*}
\end{longrotatetable}

\pagebreak % ---------------------------------------------------------------------------------------------

\pagebreak

\pagebreak

\pagebreak

\pagebreak

\pagebreak

\pagebreak

\pagebreak

\pagebreak

\pagebreak

\pagebreak

\pagebreak

\pagebreak

\pagebreak

\end{document}